\documentclass[11pt]{article}

\usepackage[preprint]{acl}

\newcommand{\paperauthors}{%
    Yuan Xu \quad Dawei Chen \quad Xiaoguang Lin \quad Xin Chen \quad Yulong Ou \quad Haijun Wu
}
\newcommand{\paperaffiliation}{Tencent}
\newcommand{\papercontact}{\texttt{xxuyyuan@foxmail.com}}

\usepackage[T1]{fontenc}
\usepackage{newtxtext,newtxmath}  
\usepackage{microtype}
\usepackage{amsmath}
\usepackage{graphicx}
\usepackage{booktabs}
\usepackage{array}
\usepackage{algorithm}
\usepackage{algpseudocode}
\usepackage{enumitem}
\usepackage{tikz}
\usetikzlibrary{arrows.meta,backgrounds,calc,decorations.pathreplacing,fit,positioning,shadows}
\usepackage{fontawesome5}
\definecolor{cBlue}{HTML}{2F6BC4}    
\definecolor{cTeal}{HTML}{118A72}    
\definecolor{cOrange}{HTML}{C8622C}  
\definecolor{cGray}{HTML}{8A96A0}    
\definecolor{cInk}{HTML}{1E2B37}
\definecolor{cL1}{HTML}{2F6BC4}      
\definecolor{cL2}{HTML}{5A8FD8}
\definecolor{cL3}{HTML}{93B6E8}

\tikzset{
  >={Stealth[length=4.2pt,width=3.4pt]},
  every picture/.append style={line cap=round, line join=round},
  module/.style={draw=#1!75!black, fill=#1!7, rounded corners=3pt, line width=0.55pt,
                 minimum height=0.9cm, align=center, inner sep=3pt, font=\small},
  module/.default=cBlue,
  frozen/.style={module=cGray, fill=cGray!12, draw=cGray!90!black},
  newmod/.style={module=cOrange, fill=cOrange!8, draw=cOrange!85!black},
  group/.style={draw=cGray!55, fill=cGray!6, rounded corners=5pt, line width=0.4pt},
  flow/.style={->, line width=0.65pt, draw=cInk!85},
  loss/.style={densely dashed, line width=0.6pt, draw=cOrange},
  losspill/.style={draw=cOrange, fill=white, rounded corners=6pt, inner sep=2.2pt, font=\footnotesize,
                   text=cOrange!80!black},
  note/.style={font=\scriptsize, text=cInk!70, align=center},
  tok/.style={draw=#1!80!black, fill=#1!22, rounded corners=2pt, minimum size=0.46cm, inner sep=0pt,
              font=\small},
  addnode/.style={circle, draw=cInk!85, line width=0.55pt, minimum size=0.34cm, inner sep=0pt, fill=white},
  pics/vec/.style={code={
    \foreach \i/\s in {0/42,1/18,2/60,3/28,4/48}
      \fill[#1!\s] (-0.11,{-0.45+0.18*\i}) rectangle ++(0.22,0.18);
    \draw[#1!80!black, line width=0.45pt] (-0.11,-0.45) rectangle (0.11,0.45);
  }},
  pics/codebook/.style={code={
    \foreach \x in {0,1,2} \foreach \y in {0,1,2,3}
      \draw[cGray!80, fill=white, line width=0.35pt] ({-0.27+0.18*\x},{-0.36+0.18*\y}) rectangle ++(0.18,0.18);
    \fill[#1, draw=#1!80!black, line width=0.35pt] (-0.09,0.0) rectangle ++(0.18,0.18);
  }},
}

\makeatletter
\let\care@thebibliography\thebibliography
\def\thebibliography#1{\care@thebibliography{#1}\setlength{\itemsep}{2pt plus 1pt}}
\makeatother
\hypersetup{pdftitle={CARE: Contrastive-Anchored Residual Encoding with Listwise Reconstruction for Retrieval-Oriented Semantic IDs},pdfauthor={Yuan Xu, Dawei Chen, Xiaoguang Lin, Xin Chen, Yulong Ou, Haijun Wu}}

\DeclareMathOperator{\vnorm}{norm}
\DeclareMathOperator*{\argmax}{arg\,max}
\DeclareMathOperator*{\argmin}{arg\,min}
\DeclareMathOperator{\softmax}{softmax}
\DeclareMathOperator{\KL}{KL}
\DeclareMathOperator{\ReLU}{ReLU}
\DeclareMathOperator{\GELU}{GELU}
\newcommand{\Lrec}{\mathcal{L}_{\mathrm{rec}}}
\newcommand{\Lanc}{\mathcal{L}_{\mathrm{anchor}}}
\newcommand{\Lnce}{\mathcal{L}_{\mathrm{NCE}}}
\newcommand{\Lsc}{\mathcal{L}_{\mathrm{cons}}}
\newcommand{\Lcl}{\mathcal{L}_{\mathrm{cl}}}
\newcommand{\Lkl}{\mathcal{L}_{\mathrm{KL}}}
\newcommand{\dhat}{\hat{d}}
\newcommand{\view}[1]{\textsf{#1}}

\setlist{itemsep=2pt,topsep=4pt}
\title{CARE: Contrastive-Anchored Residual Encoding with Listwise Reconstruction for Retrieval-Oriented Semantic IDs}
\author{\paperauthors \\[3pt] \paperaffiliation \\[1pt] {\normalsize\papercontact}}

\begin{document}
\maketitle

\begin{abstract}
Generative retrieval replaces nearest-neighbor search with the generation of a document's semantic ID (SID), so the SID tokenizer bounds what the retriever can recall. Common SID constructions (residual $k$-means, RQ-VAE, finite scalar quantization, capacity balancing) optimize criteria defined on document embeddings alone and do not control whether a query can still find its relevant documents after quantization. We propose two retrieval-oriented SID tokenizers. \textbf{CARE-SID} (Contrastive-Anchored Residual Encoding for Semantic IDs) quantizes in a retrieval-aligned space: a zero-initialized residual alignment network is trained with a query--document contrastive loss before quantization, an anchor loss and a reconstruction target fixed to the original embedding prevent drift and collapse, and reference-vector rating residual quantization assigns the codes. \textbf{CARE-LR} (CARE with Listwise Reconstruction) then distills, for each training query, the similarity distribution that the original embeddings induce over a candidate list into the SID reconstructions, through a function-preserving layer-resolved head. Both keep a document-only encoding interface. We judge SIDs without training a generator, by row-budget recall in a bucketing view (the partition) and a reconstruction view (the SID with its reconstruction). In offline evaluation on industrial video search data, both tokenizers achieve higher reconstruction-view recall than residual $k$-means, and CARE-LR improves further on CARE-SID, mainly through the listwise distillation. A comparison of two document embeddings shows that contrastive alignment helps most when the embedding has not been trained on search data, and that the embedding affects SID recall far more than the quantizer. CARE-SID is deployed in the generative retrieval stage of an industrial video search system.
\end{abstract}

\section{Introduction}
\label{sec:intro}

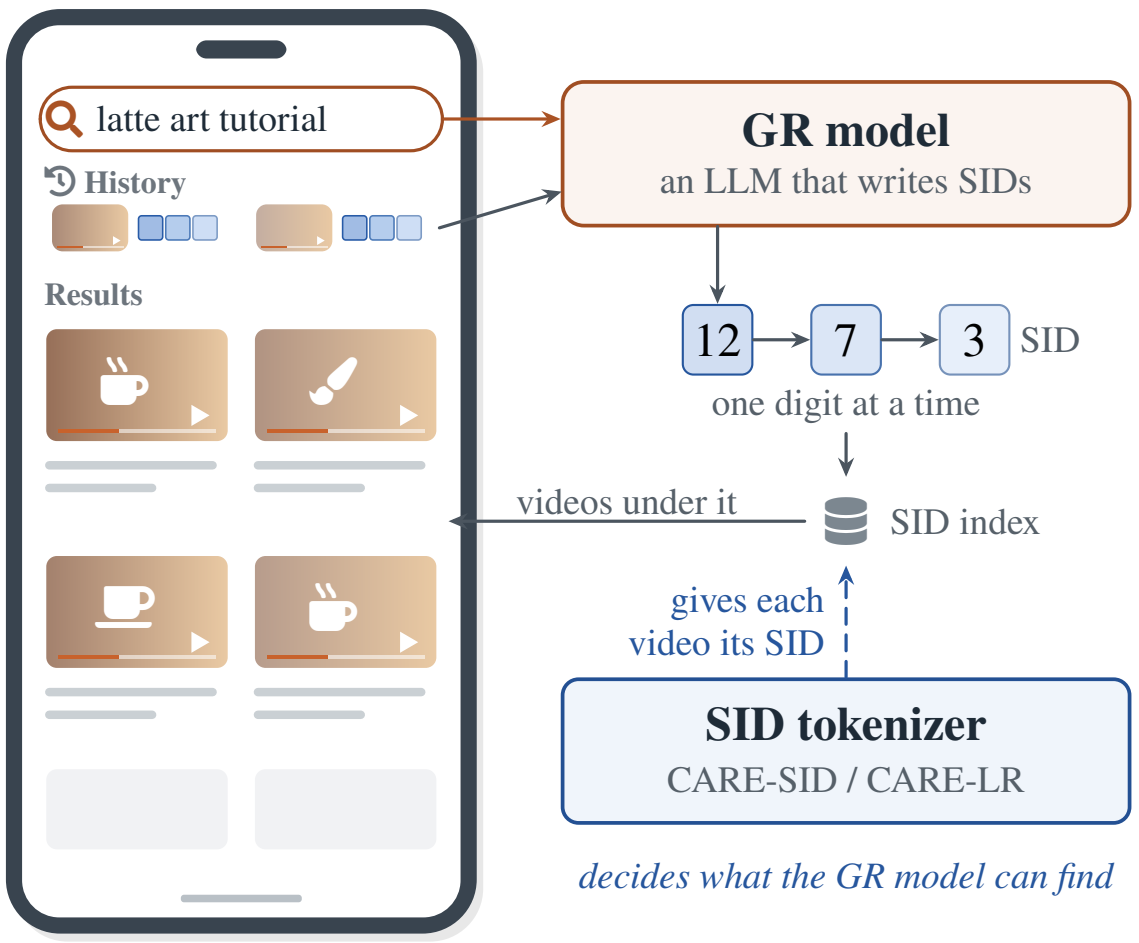
\begin{figure}[t]
  \centering
\definecolor{cCoffee}{HTML}{7B4A2E}
\definecolor{cCream}{HTML}{E9C9A3}
\begin{tikzpicture}[x=1cm, y=1cm,
    pics/thumb/.style args={#1/#2/#3}{code={%
      \begin{scope}[scale=#3]
        \shade[rounded corners=2pt, left color=cCoffee!#1, right color=cCream] (-0.6,-0.37) rectangle (0.6,0.37);
        \fill[white] (0.36,-0.27) -- (0.36,-0.13) -- (0.48,-0.2) -- cycle;
        \fill[white, opacity=0.55] (-0.52,-0.31) rectangle (0.52,-0.29);
        \fill[cOrange] (-0.52,-0.31) rectangle (-0.12,-0.29);
      \end{scope}
      \node[text=white, font=\small] at (-0.08,0.03) {#2};}},
    titleline/.style={cGray!45, line width=1.6pt, line cap=round},
    sidchip/.style={fill=#1!45, draw=#1!85!black, line width=0.3pt, rounded corners=0.6pt, minimum size=0.16cm,
                    inner sep=0pt},
    lab/.style={font=\scriptsize, text=cInk!75, inner sep=1pt},
    head/.style={font=\tiny\bfseries, text=cInk!65, anchor=west, inner sep=0pt},
    arr/.style={-{Stealth[length=3.6pt,width=3pt]}, line width=0.6pt, draw=cInk!80},
  ]
  \filldraw[cInk!88, rounded corners=10pt, drop shadow={opacity=0.18, shadow xshift=0.05cm, shadow yshift=-0.06cm}]
    (0,0) rectangle (3.1,6.1);
  \fill[white, rounded corners=7pt] (0.1,0.1) rectangle (3.0,6.0);
  \fill[cInk!88, rounded corners=2pt] (1.25,5.78) rectangle (1.85,5.9);                 
  \fill[cGray!60, rounded corners=0.6pt] (1.15,0.2) rectangle (1.95,0.25);              
  \node[draw=cOrange!80!black, fill=white, rounded corners=0.21cm, line width=0.6pt, minimum width=2.66cm,
        minimum height=0.42cm] (box) at (1.55,5.38) {};
  \node[font=\scriptsize, text=cOrange!85!black] at (0.38,5.38) {\faSearch};
  \node[lab, anchor=west, text=cInk] at (0.55,5.38) {latte art tutorial};
  \node[head] at (0.24,4.96) {\faHistory\ History};
  \foreach \x/\a in {0.55/65, 1.9/45} {
    \pic at (\x,4.66) {thumb={\a//0.42}};
    \node[sidchip=cL1] at (\x+0.4,4.66) {};
    \node[sidchip=cL2] at (\x+0.58,4.66) {};
    \node[sidchip=cL3] at (\x+0.76,4.66) {};
  }
  \node[head] at (0.24,4.22) {Results};
  \foreach \x/\y/\a/\i in {0.86/3.62/80/\faMugHot, 2.24/3.62/60/\faPaintBrush, 0.86/2.12/70/\faCoffee, 2.24/2.12/55/\faMugHot} {
    \pic at (\x,\y) {thumb={\a/\i/1}};
    \draw[titleline] (\x-0.58,\y-0.53) -- (\x+0.5,\y-0.53);
    \draw[titleline] (\x-0.58,\y-0.68) -- (\x+0.1,\y-0.68);
  }
  \foreach \x in {0.86, 2.24} \fill[cGray!12, rounded corners=2pt] (\x-0.6,0.55) rectangle (\x+0.6,1.08);

  \node[draw=cOrange!80!black, fill=cOrange!7, rounded corners=4pt, line width=0.7pt, minimum width=3.75cm,
        minimum height=0.95cm] (gr) at (5.55,5.15) {};
  \node[font=\small\bfseries, text=cInk] at (5.55,5.31) {GR model};
  \node[lab] at (5.55,4.97) {an LLM that writes SIDs};
  \draw[arr, draw=cOrange!85!black] (box.east) -- (gr.west |- box.east);
  \draw[arr] (2.86,4.66) -- (3.67,4.9);
  \foreach \x/\c/\v [count=\k] in {4.7/cL1/12, 5.55/cL2/7, 6.4/cL3/3} \node[tok=\c] (s\k) at (\x,3.92) {\v};
  \draw[arr] (4.7,4.67) -- (s1.north);
  \draw[arr] (s1.east) -- (s2.west);
  \draw[arr] (s2.east) -- (s3.west);
  \node[lab, anchor=west] at (6.66,3.92) {SID};
  \node[lab] at (5.55,3.48) {one digit at a time};
  \node[font=\small, text=cGray!90!black] (db) at (5.55,2.72) {\faDatabase};
  \node[lab, anchor=west] at (5.8,2.72) {SID index};
  \draw[arr] (5.55,3.3) -- (db.north);
  \draw[arr] (db.west) -- node[lab, above] {videos under it} (2.92,2.72);
  \node[module, minimum width=3.75cm, minimum height=0.95cm, line width=0.7pt] (tok) at (5.55,1.2) {};
  \node[font=\small\bfseries, text=cInk] at (5.55,1.38) {SID tokenizer};
  \node[lab] at (5.55,1.01) {CARE-SID / CARE-LR};
  \draw[arr, densely dashed, draw=cBlue!80!black] (tok.north) -- (db.south);
  \node[lab, text=cBlue!80!black, align=right, anchor=east] at (5.45,2.06) {gives each\\video its SID};
  \node[lab, text=cBlue!80!black, font=\scriptsize\itshape] at (5.55,0.35) {decides what the GR model can find};
\end{tikzpicture}
  \caption{Generative retrieval in video search. The GR model reads the query and the SIDs of the videos the user clicked before, and writes the semantic ID (SID) of a target video one digit at a time; the videos stored under that SID are returned. The SID tokenizer gives every video its SID and therefore decides what the GR model can find. This paper proposes two such tokenizers, CARE-SID and CARE-LR. Digit values and videos are illustrative.}
  \label{fig:teaser}
\end{figure}

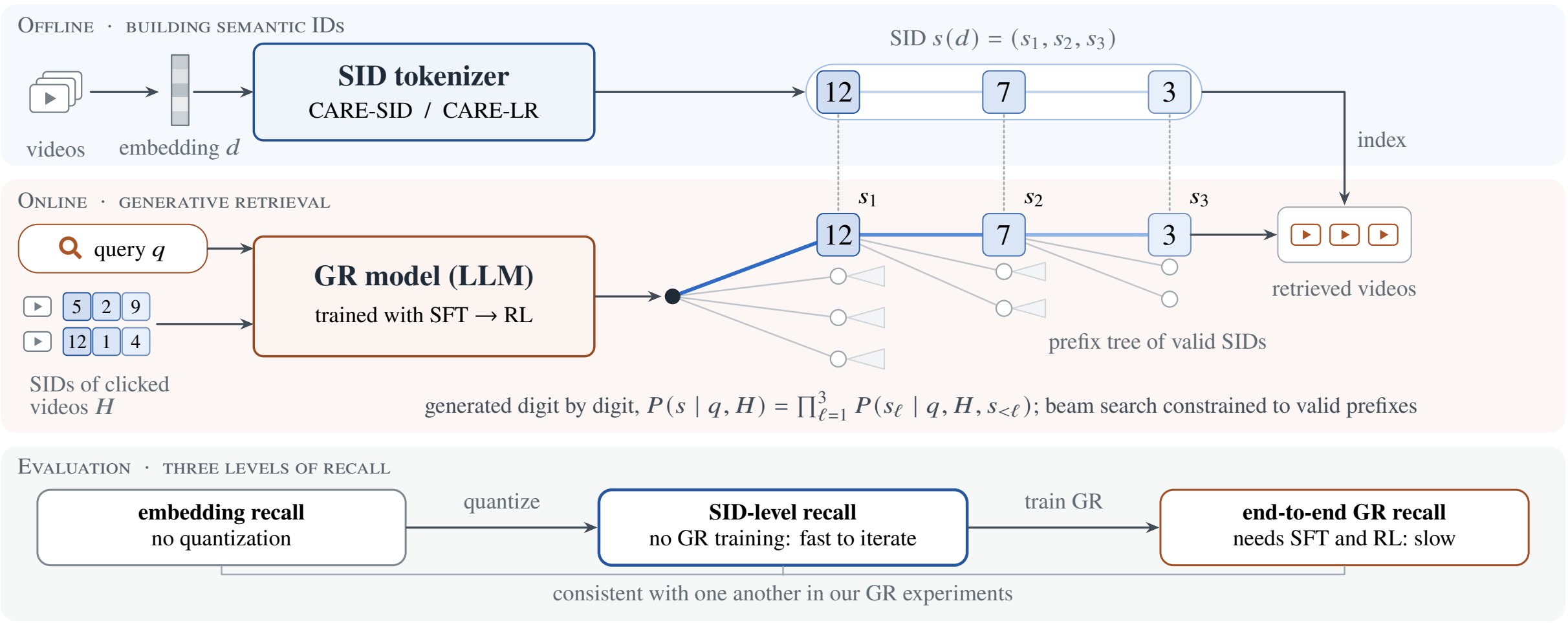
\begin{figure*}[t]
  \centering
  \resizebox{\textwidth}{!}{
\begin{tikzpicture}[x=1cm, y=1cm,
    pics/video/.style={code={
      \draw[#1!85!black, fill=white, line width=0.45pt, rounded corners=1.2pt] (-0.21,-0.15) rectangle (0.21,0.15);
      \fill[#1!85!black] (-0.045,-0.07) -- (-0.045,0.07) -- (0.075,0) -- cycle;}},
    pics/video/.default=cGray,
    band/.style={fill=#1, rounded corners=5pt},
    bandlabel/.style={font=\scriptsize\scshape, text=cInk!75, anchor=north west, inner sep=0pt},
    title/.style={font=\small\bfseries, text=cInk},
    tnode/.style={circle, fill=white, draw=cGray!85, line width=0.45pt, minimum size=0.17cm, inner sep=0pt},
    tedge/.style={draw=cGray!55, line width=0.5pt},
    pics/subtree/.style={code={\filldraw[fill=cGray!10, draw=cGray!40, line width=0.4pt, line join=round]
      (0.1,0) -- (0.5,0.11) -- (0.5,-0.11) -- cycle;}},
    hpath/.style={draw=#1, line width=1.15pt},
    guide/.style={draw=cGray!75, densely dotted, line width=0.6pt},
    chip/.style={draw=#1!80!black, fill=#1!22, rounded corners=1.2pt, minimum size=0.3cm, inner sep=0pt,
                 font=\fontsize{6}{6}\selectfont},
    evalbox/.style={draw=#1, fill=white, rounded corners=4pt, minimum width=4.0cm, minimum height=0.78cm,
                    align=center, font=\scriptsize, line width=0.6pt},
    arrowlabel/.style={note, above, text height=1.6ex, text depth=0.4ex},
  ]
  \fill[band=cBlue!5]   (0,4.35)  rectangle (17,6.1);
  \fill[band=cOrange!5] (0,1.45)  rectangle (17,4.2);
  \fill[band=cGray!8]   (0,-0.6) rectangle (17,1.3);
  \node[bandlabel] at (0.18,5.97) {Offline \;$\cdot$\; building semantic IDs};
  \node[bandlabel] at (0.18,4.07) {Online \;$\cdot$\; generative retrieval};
  \node[bandlabel] at (0.18,1.17) {Evaluation \;$\cdot$\; three levels of recall};

  \foreach \o in {2,1,0} \pic at ({0.53+0.07*\o},{5.08+0.07*\o}) {video};
  \node[note] at (0.6,4.53) {videos};
  \draw[flow] (0.98,5.15) -- (1.72,5.15);
  \pic[scale=0.85] at (1.95,5.17) {vec=cGray};
  \node[note] at (1.95,4.53) {embedding $d$};
  \draw[flow] (2.1,5.15) -- (2.75,5.15);
  \node[module, minimum width=3.7cm, minimum height=1.05cm, line width=0.7pt] (tok) at (4.6,5.15) {};
  \node[title] at (4.6,5.33) {SID tokenizer};
  \node[font=\scriptsize] at (4.6,4.95) {CARE-SID \;/\; CARE-LR};
  \node[draw=cBlue!45, fill=white, rounded corners=0.3cm, minimum width=4.3cm, minimum height=0.6cm] (sid) at (10.9,5.15) {};
  \draw[cBlue!30, line width=0.8pt] (9.1,5.15) -- (12.7,5.15);
  \foreach \x/\c/\v in {9.1/cL1/12, 10.9/cL2/7, 12.7/cL3/3} \node[tok=\c] at (\x,5.15) {\v};
  \node[note, anchor=south] at (10.9,5.47) {SID $s(d)=(s_1,s_2,s_3)$};
  \draw[flow] (tok.east) -- (sid.west);
  \draw[flow] (sid.east) -- (14.6,5.15) -- node[right, note, pos=0.42] {index} (14.6,3.92);

  \node[draw=cOrange!80!black, fill=white, rounded corners=0.21cm, minimum width=2.05cm, minimum height=0.42cm,
        font=\scriptsize] (q) at (1.22,3.45) {\textcolor{cOrange!85!black}{\faSearch}\;\;query $q$};
  \foreach \y/\a/\b/\z in {2.82/5/2/9, 2.44/12/1/4} {
    \pic[scale=0.7] at (0.4,\y) {video};
    \node[chip=cL1] at (0.83,\y) {\a}; \node[chip=cL2] at (1.15,\y) {\b}; \node[chip=cL3] at (1.47,\y) {\z};
  }
  \node[note, anchor=north west, align=left] at (0.18,2.2) {SIDs of clicked\\[-1pt]videos $H$};
  \draw[flow] (q.east) -- (2.75,3.45);
  \draw[flow] (1.7,2.63) -- (2.75,2.63);
  \node[module=cOrange, minimum width=3.7cm, minimum height=1.3cm, line width=0.7pt] (gr) at (4.6,2.93) {};
  \node[title] at (4.6,3.13) {GR model (LLM)};
  \node[font=\scriptsize] at (4.6,2.73) {trained with SFT $\rightarrow$ RL};

  \coordinate (root) at (7.3,2.93);
  \foreach \y in {3.15,2.7,2.25} {
    \draw[tedge] (root) -- (9.1,\y);
    \pic at (9.1,\y) {subtree};
    \node[tnode] at (9.1,\y) {};
  }
  \foreach \y in {3.2,2.8} {
    \draw[tedge] (9.1,3.6) -- (10.9,\y);
    \pic[scale=0.9] at (10.9,\y) {subtree};
    \node[tnode] at (10.9,\y) {};
  }
  \foreach \y in {3.25,2.9} { \draw[tedge] (10.9,3.6) -- (12.7,\y); \node[tnode] at (12.7,\y) {}; }
  \draw[hpath=cL1] (root) -- (9.1,3.6);
  \draw[hpath=cL2] (9.1,3.6) -- (10.9,3.6);
  \draw[hpath=cL3] (10.9,3.6) -- (12.7,3.6);
  \fill[cInk] (root) circle (0.085);
  \node[note, anchor=west] at (11.25,2.42) {prefix tree of valid SIDs};
  \draw[flow] (gr.east) -- ([xshift=-0.12cm]root);
  \foreach \x/\c/\v/\l in {9.1/cL1/12/1, 10.9/cL2/7/2, 12.7/cL3/3/3} {
    \draw[guide] (\x,4.9) -- (\x,3.84);
    \node[tok=\c] at (\x,3.6) {\v};
    \node[font=\scriptsize, anchor=west] at (\x+0.08,3.98) {$s_\l$};
  }
  \node[draw=cGray!80, fill=white, rounded corners=3pt, minimum width=1.45cm, minimum height=0.6cm] (bucket) at (14.6,3.6) {};
  \foreach \x in {-0.42,0,0.42} \pic[scale=0.75] at (14.6+\x,3.6) {video=cOrange};
  \draw[flow] (12.93,3.6) -- (bucket.west);
  \node[note, anchor=north] at (14.6,3.24) {retrieved videos};
  \node[font=\scriptsize, text=cInk!85] at (10.0,1.74)
    {generated digit by digit, $P(s\mid q,H)=\prod_{\ell=1}^{3}P(s_\ell\mid q,H,s_{<\ell})$; beam search constrained to valid prefixes};

  \node[evalbox=cGray!90!black] (e) at (2.4,0.42) {\textbf{embedding recall}\strut\\no quantization\strut};
  \node[evalbox=cBlue!80!black, line width=0.9pt] (s) at (8.5,0.42)
    {\textbf{SID-level recall}\strut\\no GR training: fast to iterate\strut};
  \node[evalbox=cOrange!80!black] (g) at (14.6,0.42) {\textbf{end-to-end GR recall}\strut\\needs SFT and RL: slow\strut};
  \draw[flow] (e.east) -- node[arrowlabel] {quantize} (s.west);
  \draw[flow] (s.east) -- node[arrowlabel] {train GR} (g.west);
  \draw[cGray!80, line width=0.5pt] (e.south) -- ++(0,-0.09) -| (g.south) (s.south) -- ++(0,-0.09);
  \node[note, anchor=north, inner sep=0pt] at (8.5,-0.2) {consistent with one another in our GR experiments};
\end{tikzpicture}\unskip}
  \caption{Semantic IDs in generative retrieval for search. Offline (top): an SID tokenizer (CARE-SID or CARE-LR in this paper) encodes the embedding $d$ of every video into a three-digit SID $s(d)$, a path in the prefix tree of valid SIDs under which the video is indexed. Online (middle): the GR model reads the query and the SIDs of previously clicked videos and generates an SID digit by digit, with beam search constrained to valid prefixes; the videos indexed under that SID are retrieved. The GR model is trained with supervised fine-tuning (SFT) followed by reinforcement learning (RL). Evaluation (bottom): recall can be measured on the embeddings, on the SIDs or on the outputs of the GR model; this paper reports SID-level recall, which needs no GR training (Section~\ref{sec:setup}).}
  \label{fig:overview}
\end{figure*}

Search systems usually retrieve candidates by embedding queries and documents into a shared space and running approximate nearest-neighbor search~\citep{karpukhin2020dpr,malkov2020hnsw}. Generative retrieval (GR) takes a different route (Figure~\ref{fig:teaser}). Every document receives a short identifier made of discrete tokens, and a sequence model reads the query and generates the identifier of a relevant document token by token~\citep{tay2022dsi,wang2022nci,sun2023genret}. When the identifiers are derived from document content, they are called semantic IDs (SIDs)~\citep{rajput2023tiger}. SIDs are now part of industrial search and recommendation systems~\citep{deng2025onerec,chen2025onesearch,chen2025unisearch,zhang2025gpr,zhu2026cqsid,han2026wegr} and are also used as learnable discrete features~\citep{xia2026qarmv2}.

An SID is the address at which the generator has to find a document (Figure~\ref{fig:overview}). Documents that share an SID can only be retrieved together, and the prefixes of the SIDs form the coarse-to-fine space that the generator walks through. Whatever query--document relevance the upstream embedding has learned, the part that quantization discards is no longer available to the generator, which sees documents only through their SIDs. What the SIDs preserve therefore bounds what generative retrieval can recall.

We first reproduced the main families of SID construction on the same search corpus: residual $k$-means (RQ-KMeans), RQ-VAE~\citep{lee2022rqvae,rajput2023tiger}, a hybrid of residual $k$-means and finite scalar quantization (Res-KmeansFSQ)~\citep{mentzer2024fsq,xia2026qarmv2}, and post-hoc capacity balancing of the last level. Each family optimizes a criterion defined on documents alone: geometric distortion, reconstruction error, fit to a lattice, or bucket capacity. Search, however, depends on relations between queries and documents. Quantization always discards information, and what matters is where the loss falls. If it falls on the directions that a query uses to separate relevant from irrelevant documents, retrieval degrades even when the distortion is small: two reconstructions with the same per-document error can preserve or swap the order of a query's top candidates (Figure~\ref{fig:kl}). An SID tokenizer for search should therefore be built and judged by query--document retrieval, which raises two questions: in which space to quantize, so that the loss avoids the directions that queries rely on, and how to keep each query's ranking of its candidates once documents are quantized.

Recent search systems have begun to bring query--item signals into SID construction through contrastive objectives on the representation before quantization~\citep{zhu2026cqsid,han2026wegr,chen2025unisearch}. We propose two tokenizers, one for each question. \textbf{CARE-SID} addresses \emph{where to quantize}. It places a residual alignment network in front of the rating residual quantizer of R3-VAE~\citep{wan2026r3vae} and trains it with a query--document contrastive loss, so that queries and their relevant documents move closer together in the space being quantized. An anchor loss and a reconstruction target fixed to the original embedding prevent drift and collapse, and a zero-initialized output layer lets training start exactly from an R3-VAE model that is itself warm-started from $k$-means codebooks. \textbf{CARE-LR} addresses \emph{how to keep the ranking after quantization}. For each training query, it distills the softmax distribution of cosine similarities over a candidate list, computed with the original embeddings, into the same distribution computed with SID reconstructions; related distillation objectives have been used to train vector quantizers for nearest-neighbor search~\citep{xiao2022distillvq} and residual quantizers for generative retrieval~\citep{chen2026pro}. The distillation passes through a layer-resolved reconstruction head that initially computes exactly the same function as the CARE-SID head. In both methods, queries enter only the training objectives; each document is encoded from its own embedding, so new documents can be encoded incrementally. Both methods also treat the original embedding as the reference for what quantization must preserve (as anchor and reconstruction target in CARE-SID, as teacher in CARE-LR), and each starts exactly from the model it extends.

Because end-to-end evaluation requires training the GR model, we evaluate SIDs directly with a row-budget recall protocol that computes the exact expected recall when a bucket is only partly consumed and that separates the quality of the bucketing from that of the reconstruction; for R3-VAE and the CARE models, the reconstruction also draws on continuous per-document coefficients that are not part of the SID (Section~\ref{sec:setup}). In offline evaluation on industrial video search data, CARE-SID raises reconstruction-view Recall@10 from 62.75\% (RQ-KMeans) to 65.23\%, and CARE-LR raises it to 66.09\%, a paired gain of 3.34 points over RQ-KMeans (95\% CI [+2.30, +4.36]). The gain of CARE-LR over CARE-SID (+0.86, [+0.55, +1.19]) comes mainly from the listwise distillation. These results use a search-tuned document embedding, on which the R3-VAE quantizer underlying CARE-SID already reaches 65.15\%; the contrastive alignment of CARE-SID adds most when the embedding has not been trained on search data, and the choice of embedding moves recall far more than the choice of quantizer (Section~\ref{sec:exp-emb}). Our contributions are:
\begin{itemize}[leftmargin=1.3em]
  \item \textbf{CARE-SID}, a tokenizer that quantizes in a learned, retrieval-aligned space: a zero-initialized residual alignment network is trained with a contrastive loss before quantization and is tied to the original embedding through both an anchor loss and the reconstruction target. Codes are assigned by reference-vector rating residual quantization, warm-started from RQ-KMeans through R3-VAE (Section~\ref{sec:care}). CARE-SID is deployed in the generative retrieval stage of an industrial video search system (Section~\ref{sec:deploy}).
  \item \textbf{CARE-LR}, a listwise reconstruction stage on top of CARE-SID that distills query-conditioned candidate distributions from the original embeddings into SID reconstructions. It starts exactly from CARE-SID through a function-preserving layer-resolved head and keeps the first-level codes and the document-only encoding interface unchanged (Section~\ref{sec:carelr}).
  \item A \textbf{retrieval-oriented evaluation protocol} for SIDs that needs no generator training, with an exact within-bucket expectation, separate bucketing and reconstruction views, and query-grouped paired bootstrap intervals (Section~\ref{sec:setup}).
  \item An \textbf{empirical study} with ablations and a comparison of two document embeddings, which locates the gain of CARE-LR in the listwise distillation and shows when contrastive alignment pays off (Section~\ref{sec:exp}).
\end{itemize}

\section{Related Work}
\label{sec:related}

\paragraph{Generative retrieval and semantic IDs.}
DSI~\citep{tay2022dsi} showed that a sequence-to-sequence model can map queries directly to document identifiers, NCI~\citep{wang2022nci} improved the decoder, and GenRet~\citep{sun2023genret} learned the tokenization. TIGER~\citep{rajput2023tiger} built semantic IDs with RQ-VAE and generated them for sequential recommendation. RIPOR~\citep{zeng2024ripor} argued that document identifiers should be built from query--document relevance rather than from document content alone: it quantizes document representations from a relevance-trained encoder and scales generative retrieval to millions of passages with prefix-oriented ranking optimization. Industrial systems now generate SIDs for recommendation, search and advertising~\citep{deng2025onerec,chen2025onesearch,chen2025unisearch,zhang2025gpr}. CQ-SID~\citep{zhu2026cqsid} builds semantic cluster IDs for e-commerce search from category and query--item signals, WeGR~\citep{han2026wegr} unifies behavior-aware tokenization and preference alignment for generative retrieval in search, and QARM V2~\citep{xia2026qarmv2} uses SIDs as learnable features for user sequence modeling. Our work shares RIPOR's premise but concerns the tokenizer alone, under a fixed upstream retrieval embedding and a separately trained generator, and is complementary to improvements of the generator.

\paragraph{SID tokenizers.}
Most SID tokenizers are residual quantizers~\citep{chen2010rvq}: each level quantizes what the previous levels left over, using $k$-means~\citep{lloyd1982}, learned codebooks as in VQ-VAE and RQ-VAE~\citep{vandenoord2017vqvae,lee2022rqvae}, or fixed lattices as in FSQ~\citep{mentzer2024fsq}. R3-VAE~\citep{wan2026r3vae} removes a learned reference direction and rates codewords by their cosine with the residual; CARE-SID uses this rating residual quantization, in the top-1 form of the public R3-VAE implementation, as its kernel. PRQ-KMeans~\citep{luo2026prq} brings projection residuals to residual $k$-means, and RQ-Kmeans+ in GPR~\citep{zhang2025gpr} initializes the codebooks of an RQ-VAE with RQ-KMeans and adds a residual connection to its encoder so that the encoder output stays close to its input early in training. LC-Rec~\citep{zheng2024lcrec} assigns conflict-free indices through uniform semantic mapping, and LETTER, CoST and SimCIT add collaborative or contrastive objectives to item tokenization~\citep{wang2024letter,zhu2024cost,zhai2025simcit}. Closest to our setting, search systems inject query--item signals into the representation before quantization: CQ-SID applies a bidirectional query--item InfoNCE loss to a trainable encoder, WeGR adds query-to-item and item-to-item contrastive objectives, and UniSearch trains its video encoder with a residual query--video contrastive loss jointly with the generator. CARE-SID shares the idea of aligning documents with queries before quantization and differs in how the alignment is attached to an existing retrieval stack: it keeps the upstream retrieval embedding frozen and learns only a residual alignment whose zero-initialized output makes training start exactly from the warm-started quantizer (a residual warm start like that of RQ-Kmeans+, but exact at initialization), and an explicit anchor loss keeps the aligned space close to the original embedding, which also remains the reconstruction target. Our design analysis found this anchoring necessary to avoid bucket merging (Section~\ref{sec:care-design}). We also measure when such alignment pays off: its gain is large for an embedding that has not been trained on search data and narrower for a search-tuned one (Section~\ref{sec:exp-emb}).

\paragraph{Retrieval-oriented quantization and distillation.}
Product quantization (PQ) and its optimized variant minimize distortion for compressed nearest-neighbor search~\citep{jegou2011pq,ge2013opq}. JPQ~\citep{zhan2021jpq} trains the query encoder and the PQ centroids jointly with a ranking loss, RepCONC~\citep{zhan2022repconc} learns discrete representations with constrained clustering, and MoPQ~\citep{xiao2021mopq} argues that quantization should target query--document matching rather than reconstruction and optimizes the matching probability of queries against reconstructed documents. Distillation transfers soft output distributions from a teacher to a student~\citep{hinton2015kd}; listwise objectives compare whole score distributions over a list~\citep{cao2007listnet}, and RocketQAv2~\citep{ren2021rocketqav2} aligns the listwise relevance distributions of a retriever and a re-ranker. Distill-VQ~\citep{xiao2022distillvq} brings this to vector quantization: the dense embeddings act as the teacher, and inverted-file (IVF) and PQ indexes are trained with listwise objectives such as KL divergence and ListNet so that the quantized embeddings reproduce the relevance the teacher assigns to documents sampled for each query, without requiring relevance labels. For semantic IDs, PRO~\citep{chen2026pro} distills the query-conditioned rankings that pre-quantization embeddings induce over code prefixes into those given by the partial reconstructions of a residual quantizer, and TopoTok~\citep{liu2026topotok} distills the item--item similarity structure of the embedding into an item tokenizer. CARE-LR shares with Distill-VQ and PRO the use of the original embedding as the teacher and differs in the student and in the lists: its student is the full reconstruction that a function-preserving layer-resolved head computes from the reference component and the rating components along the document's SID path, and its candidate lists are drawn from the stages of a real search funnel in addition to nearest neighbors and random documents. Its first-level codes stay fixed, so the top of the prefix tree is unchanged, and each document is still encoded from its own embedding.

\paragraph{Evaluating SIDs without a generator.}
Because training a generator for every candidate tokenizer is slow, several works judge SIDs directly. FORGE~\citep{fu2026forge} proposes an embedding hit rate and the Gini coefficient of SID usage, which correlate with the hit rate of a trained generator, and R3-VAE~\citep{wan2026r3vae} introduces semantic cohesion and preference discrimination; these proxies are computed without queries and describe the embedding, the clusters or the usage of codes. For document identifiers in generative retrieval, \citet{allal2026sidspaces} find that the uniqueness ratio flags collapse but that none of their intrinsic, training-free metrics explains retrieval effectiveness among identifier spaces without collapse, and \citet{zhang2026faithful} show that hit rates counted on SIDs instead of items are inflated by collisions. Our protocol (Section~\ref{sec:setup}) instead measures query--document recall through the SIDs: it counts documents rather than SIDs within a fixed budget of document rows, applies tie-aware evaluation~\citep{mcsherry2008ties} to the partly consumed bucket, and separates the quality of the partition from that of the reconstruction.

\section{Problem Setup and Retrieval-Oriented Evaluation}
\label{sec:setup}

\paragraph{Semantic IDs.}
Each document has a unit-norm embedding $d\in\mathbb{R}^{D}$ produced by an upstream retrieval model, and queries $q$ are embedded in the same space. A three-level tokenizer maps $d$ to an SID $s(d)=(s_1,s_2,s_3)$ whose $\ell$-th digit $s_\ell\in\{1,\dots,K_\ell\}$ is the code selected at level $\ell$. We use $D=256$ and $(K_1,K_2,K_3)=(16384,4096,2048)$, which gives 22,528 code tokens (each level has its own) and $2^{37}\approx1.4\times10^{11}$ possible SIDs. Code indices carry no meaning by themselves; documents that share a prefix have followed the same quantization path, so prefixes form a coarse-to-fine hierarchy. We require \emph{document-only encoding}: the SID of a document is computed from its own embedding, without queries and without the rest of the corpus, so that new documents can be encoded incrementally. A tokenizer may also produce a continuous reconstruction $\dhat$ of $d$. To monitor collisions, we use $\mathrm{ICR}=\sum_b\mathbf{1}[n_b=1]\big/\sum_b n_b$, the fraction of documents that occupy a full SID alone, where $n_b$ is the number of documents in bucket $b$. ICR detects codebook collapse, but it does not measure retrieval: random IDs over the $2^{37}$ SIDs reach an ICR close to 1 and carry no semantics.

\paragraph{Why evaluate at the SID level.}
The quality of an SID is ultimately judged end to end, by the recall of the generative retrieval (GR) model that uses it. In our search setting, the GR model reads the query and the SIDs of the documents the user clicked before, and generates the SID of the document to retrieve (Section~\ref{sec:deploy}). Training it involves supervised fine-tuning followed by reinforcement learning, so every end-to-end comparison is slow. We therefore first judge SIDs with SID-level recall, computed from the SID buckets and the SID reconstructions (Figure~\ref{fig:overview}, bottom): it needs no GR training and lets us iterate on the tokenizer quickly. In our subsequent GR experiments, the three recall measures (on the embeddings, on the SIDs and on the outputs of the GR model) were consistent with one another; in particular, GR recall ranked CARE-SID above RQ-KMeans, as the reconstruction view does, which supports the SID-level metric as an intermediate signal. End-to-end GR results will be added in a later version of this paper.

\paragraph{Row-budget recall.}
Each query $q$ comes with a set $G(q)$ of target documents, which we call its ideal documents (Section~\ref{sec:exp} describes how they were obtained). A retrieval view ranks buckets by the cosine similarity between $q$ and a bucket key and then takes documents bucket by bucket until a budget of $N$ document rows is spent. The last bucket is usually consumed only in part. All documents in a bucket share one score, so instead of a random draw we compute the exact expectation under uniform sampling without replacement inside that bucket, as in tie-aware evaluation~\citep{mcsherry2008ties}. If the bucket holds $n$ rows, $m\le n$ rows of budget remain when it is reached, and an ideal document has $c$ copies in it, the document is retrieved with probability
\begin{equation}
  \pi \;=\; 1-\binom{n-c}{m}\Big/\binom{n}{m},
  \label{eq:pi}
\end{equation}
which reduces to $m/n$ for $c=1$; for example, a document that appears once in a bucket of 100 rows is retrieved with probability 0.1 when 10 rows of budget remain. When an ideal document appears in several buckets, the probabilities of missing it multiply. Recall@$N$ of a query is the expected fraction of $G(q)$ retrieved within $N$ rows; ideal documents that are missing from the evaluation pool still count in the denominator. We report the mean over queries in percent, write @$N$ for Recall@$N$ when the view is clear, and give differences between methods in percentage points (points).

\paragraph{Retrieval views.}
The views in Table~\ref{tab:views} differ only in their buckets and bucket keys. The \view{raw\_dense} view ranks every row by $\cos(q,d)$ and serves as an unquantized reference. The \emph{bucketing view} \view{rawmean\_P$\ell$} groups documents by their first $\ell$ code digits and keys each bucket by the mean of its original embeddings, so it measures the partition alone. The \emph{reconstruction view} \view{recon\_P3} groups documents by full SID and keys each bucket by the mean of its reconstructions, so it measures how much retrieval ability the SID and its reconstruction pathway retain together; \view{recon\_doc} ranks the reconstructions row by row. We fixed \view{recon\_P3}@10 as the primary metric before the CARE-LR experiments; the other cutoffs and views are secondary. In our pool a first-level bucket holds about 61 rows on average, so at $N=10$ the one-digit view cannot consume even a typical bucket, while under full SIDs about 98\% of rows lie in singleton buckets, so that \view{rawmean\_P3} is nearly identical to \view{raw\_dense} and \view{recon\_P3} to \view{recon\_doc}. We therefore read the bucketing view at one- and two-digit prefixes and the reconstruction view at full SIDs. For R3-VAE and the CARE models, the reconstruction uses the continuous rating and reference coefficients of each document, which are not part of the integer SID, whereas the RQ-KMeans reconstruction is the sum of the selected codewords. All methods share the retrieval protocol, but not the amount of information available to the reconstruction.

\begin{table*}[t]
\centering
\small
\caption{Retrieval views. All views rank buckets by $\cos(q,\text{key})$ and spend a budget of $N$ document rows.}
\label{tab:views}
\begin{tabular}{llll}
\toprule
View & Buckets & Bucket key & What it measures \\
\midrule
\view{raw\_dense} & every row & original embedding $d$ & unquantized reference \\
\view{rawmean\_P$\ell$} & first $\ell$ code digits & mean of original embeddings & quality of the partition \\
\view{recon\_P3} & full SID & mean of reconstructions $\dhat$ & SID together with its reconstruction \\
\view{recon\_doc} & every row & reconstruction $\dhat$ & reconstruction alone \\
\bottomrule
\end{tabular}
\end{table*}

\paragraph{Statistics.}
Confidence intervals for differences between methods come from a paired bootstrap with 2,000 resamples over query texts~\citep{efron1993bootstrap}; a text in several query segments is resampled as one cluster, and we report 95\% percentile intervals, which are conditional on fixed checkpoints and a fixed pool (see Limitations).

\section{Revisiting Classic SID Construction}
\label{sec:classic}

Before introducing our methods, we revisit the four families of SID construction that we reproduced on the same search document embeddings (Table~\ref{tab:classic}). Each has a clear strength, but each optimizes a criterion that involves documents only. RQ-KMeans and R3-VAE, the RQ-VAE variant on which CARE-SID builds, also serve as the quantitative baselines in Section~\ref{sec:exp}.

\paragraph{Residual $k$-means.}
RQ-KMeans clusters residuals level by level. With $e_0=d$,
\begin{equation}
  \begin{gathered}
  s_\ell=\argmin_k \lVert e_{\ell-1}-C_\ell[k]\rVert^2,\\
  e_\ell=e_{\ell-1}-C_\ell[s_\ell],\quad \dhat=\textstyle\sum_{\ell=1}^{3}C_\ell[s_\ell].
  \end{gathered}
\end{equation}
We fit three global codebooks with distributed $k$-means over the corpus; the second- and third-level codebooks are shared by all documents rather than trained per parent cluster. The method is simple, reproducible and scalable to production corpora, and it initializes the learned methods below. Its limitation lies in the objective: minimizing the mean distortion is not the same as maximizing recall, and dense regions of the embedding distribution turn into hot buckets. Projection variants such as PRQ-KMeans~\citep{luo2026prq} give more balanced buckets, but in our runs more balanced bucketing did not by itself improve retrieval, in line with a reproducibility study of SID designs for generative recommendation~\citep{chen2026goodsid}.

\paragraph{RQ-VAE.}
RQ-VAE~\citep{lee2022rqvae,rajput2023tiger} places a learned encoder before the residual codebooks and a decoder after them and trains with reconstruction, codebook and commitment losses. The encoding space and the codebooks become learnable, at the cost of low codebook utilization, representation collapse and sensitivity to initialization. More fundamentally, the reconstruction loss asks whether a document can be restored, not whether it should rank above another document for some query. R3-VAE~\citep{wan2026r3vae} adds a reference vector and rating residual quantization and trains more stably; CARE-SID builds on this kernel.

\paragraph{Res-KmeansFSQ.}
Following QARM V2~\citep{xia2026qarmv2}, we keep $k$-means for the first two levels and replace the third with finite scalar quantization~\citep{mentzer2024fsq}: the residual $e_2$ left after the first two levels is projected to 13 dimensions, squashed by a sigmoid and rounded per dimension, $b=\operatorname{round}(\operatorname{sigmoid}(W e_2))\in\{0,1\}^{13}$, which yields $2^{13}=8192$ codes without learned centroids. A regular lattice does not imply uniform code usage. Each bit is close to balanced on its own, but the bits are correlated, so the combined codes remain unevenly used. In our reproduction, an entropy term on code usage is computed on the hard rounded codes and sends no gradient to the projection, so it can only monitor the imbalance.

\paragraph{Third-level capacity balancing.}
Capacity balancing post-processes existing SIDs. It fixes the first two digits and reassigns the third within each two-digit prefix according to a target mean bucket size and a hard cap. Our implementation proposes to each document the third-level centers nearest to its original center, accepts documents in two capacity-limited rounds, fills empty slots with the remaining documents and orders documents within a bucket by a hash of their ID. This suppresses hot buckets, but it moves some documents away from their nearest center; consistent with this, WeGR~\citep{han2026wegr} finds that third-level balancing raises codebook utilization but leaves the hit rate of generative retrieval relatively low. Moreover, because the assignment depends on the whole batch, a new document cannot reproduce it from the codebooks alone.

\begin{table*}[t]
\centering
\small
\caption{Classic SID constructions reproduced on the same search corpus.}
\label{tab:classic}
\renewcommand{\arraystretch}{1.15}
\begin{tabular}{>{\raggedright}p{2.3cm}>{\raggedright}p{3.3cm}>{\raggedright}p{3.7cm}>{\raggedright\arraybackslash}p{5.0cm}}
\toprule
Method & Objective & Strength & Limitation for retrieval \\
\midrule
RQ-KMeans & residual distortion, level by level & simple, reproducible, distributed training at corpus scale & ignores queries; dense regions become hot buckets \\
\addlinespace[2pt]
RQ-VAE & reconstruction + codebook and commitment losses & learnable encoding space and codebooks & collapse-prone and sensitive to initialization; ignores queries \\
\addlinespace[2pt]
\mbox{Res-KmeansFSQ} & fit to a fixed lattice at level 3 & fixed capacity, no learned centroids & a uniform lattice does not give uniform code usage \\
\addlinespace[2pt]
Capacity balancing & bucket capacity (post-hoc reassignment) & bounds hot buckets & moves documents off their nearest center; not reproducible per document \\
\bottomrule
\end{tabular}
\end{table*}

\paragraph{Common limitation.}
All four criteria are defined on document embeddings alone: geometric distortion, reconstruction error, lattice fit or bucket capacity. Retrieval, in contrast, depends on how a query ranks documents relative to each other. This leads to the two questions behind our methods. CARE-SID asks in which space to quantize, and CARE-LR asks how to preserve the retrieval ranking once documents are quantized.

\section{CARE-SID: Quantizing in a Retrieval-Aligned Space}
\label{sec:care}

\begin{figure*}[t]
  \centering
  \resizebox{\textwidth}{!}{
\begin{tikzpicture}[x=1cm, y=1cm]
  \pic at (0.3,0) {vec=cGray};
  \node[note, anchor=north west] at (0.0,-0.5) {document $d$};
  \node[module, minimum width=1.25cm] (f) at (1.75,0) {$f$\\[-1pt]{\scriptsize MLP}};
  \node[addnode] (add) at (3.0,0) {};
  \draw[line width=0.6pt] (add.north) -- (add.south) (add.west) -- (add.east);
  \draw[flow] (0.42,0) -- (f.west);
  \draw[flow] (f.east) -- (add.west);
  \draw[flow] (0.8,0) -- (0.8,0.7) -| (add.north);                       
  \pic at (4.15,0) {vec=cBlue};
  \node[font=\small, anchor=west] at (4.27,0.3) {$z$};
  \draw[flow] (add.east) -- node[above, font=\scriptsize] {norm} (4.02,0);

  \node[module, minimum width=1.55cm] (ref) at (5.8,0) {$-\,(z^{\top}r)\,r$};
  \node[note, above=1pt] at (ref.north) {reference $r$};
  \draw[flow] (4.28,0) -- (ref.west);
  \begin{scope}[on background layer]
    \node[group, fit={(7.0,-0.68) (13.05,0.72)}, inner sep=0pt] {};
  \end{scope}
  \foreach \l/\x/\c in {1/7.9/cL1, 2/10.0/cL2, 3/12.1/cL3} {
    \node[module, minimum width=1.45cm, minimum height=1.0cm] (L\l) at (\x,0) {};
    \pic at ({\x-0.33},0) {codebook=\c};
    \node[font=\small] at ({\x+0.33},0) {$C_{\l}$};
    \node[tok=\c] (s\l) at (\x,1.6) {$s_{\l}$};
    \draw[flow] (L\l.north) -- (s\l.south);
  }
  \draw[flow] (ref.east) -- node[above, font=\scriptsize] {$e_0$} (L1.west);
  \draw[flow] (L1.east) -- node[above, font=\scriptsize] {$e_1$} (L2.west);
  \draw[flow] (L2.east) -- node[above, font=\scriptsize] {$e_2$} (L3.west);
  \draw[decorate, decoration={brace, amplitude=4pt}, line width=0.5pt]
    ([xshift=-2pt, yshift=3pt]s1.north west) -- ([xshift=2pt, yshift=3pt]s3.north east)
    node[midway, above=4pt, font=\small] {semantic ID $(s_1,s_2,s_3)$};

  \coordinate (bus) at (0,-1.6);
  \foreach \src/\lab in {ref/0, L1/1, L2/2, L3/3}
    \draw[line width=0.65pt, draw=cInk!85] (\src.south) -- node[right, font=\scriptsize, pos=0.5] {$v_{\lab}$} (\src.south |- bus);
  \node[addnode] (sum) at (13.3,-1.6) {};
  \draw[line width=0.6pt] (sum.north) -- (sum.south) (sum.west) -- (sum.east);
  \draw[flow] (ref.south |- bus) -- (sum.west);
  \node[module, minimum width=1.2cm] (g) at (14.5,-1.6) {$g$\\[-1pt]{\scriptsize MLP}};
  \draw[flow] (sum.east) -- (g.west);
  \pic at (15.8,-1.6) {vec=cBlue};
  \node[note, below=0.5cm] at (15.8,-1.6) {reconstruction $\hat d$};
  \draw[flow] (g.east) -- (15.67,-1.6);

  \pic at (0.3,-2.75) {vec=cOrange};
  \node[note, anchor=north west] at (0.0,-3.25) {query $q$};
  \node[module, minimum width=1.25cm, dashed] (fq) at (1.75,-2.75) {$f$\\[-1pt]{\scriptsize shared}};
  \pic at (4.15,-2.75) {vec=cOrange};
  \node[font=\small, anchor=west] at (4.27,-2.75) {$z_q$};
  \draw[flow, draw=cOrange!90!black] (0.42,-2.75) -- (fq.west);
  \draw[flow, draw=cOrange!90!black] (fq.east) -- (4.02,-2.75);

  \draw[loss] (4.15,0.47) -- (4.15,0.86);
  \node[losspill, anchor=south] at (4.15,0.86) {$\mathcal{L}_{\mathrm{anchor}}(z,d)$};
  \draw[loss] (4.15,-0.47) -- (4.15,-2.28);
  \node[losspill] at (4.15,-1.37) {$\mathcal{L}_{\mathrm{NCE}}(z_q,z)$};
  \draw[loss] (15.8,-1.13) -- (15.8,-0.62);
  \node[losspill, anchor=south] at (15.8,-0.62) {$\mathcal{L}_{\mathrm{rec}}(\hat d,d)$};

  \begin{scope}[shift={(7.4,-2.75)}, font=\scriptsize, every node/.append style={text height=1.5ex, text depth=.25ex}]
    \draw[flow] (0,0) -- (0.55,0); \node[anchor=west] at (0.6,0) {encoding (document only)};
    \draw[loss] (4.25,0) -- (4.8,0); \node[anchor=west] at (4.85,0) {training loss};
    \node[module, minimum height=0.3cm, minimum width=0.42cm, inner sep=0pt] at (7.0,0) {};
    \node[anchor=west] at (7.25,0) {trainable};
  \end{scope}
\end{tikzpicture}\unskip}
  \caption{CARE-SID. Dark solid arrows use only the document itself. Along the encoding path, the residual alignment network maps $d$ to $z=\vnorm(d+f(d))$, the reference component $v_0$ is removed, and rating residual quantization (gray panel) picks one codeword per level and emits the semantic ID $(s_1,s_2,s_3)$. The reference component and the rating components $v_1,v_2,v_3$ are summed and decoded by $g$ into the reconstruction $\hat d$. Dashed lines mark the reconstruction, anchor and contrastive losses, each attached to the vector it supervises (the consistency and co-cluster terms of Section~\ref{sec:care-loss} are not drawn); queries (orange) pass through the shared $f$ only during training.}
  \label{fig:care-sid}
\end{figure*}
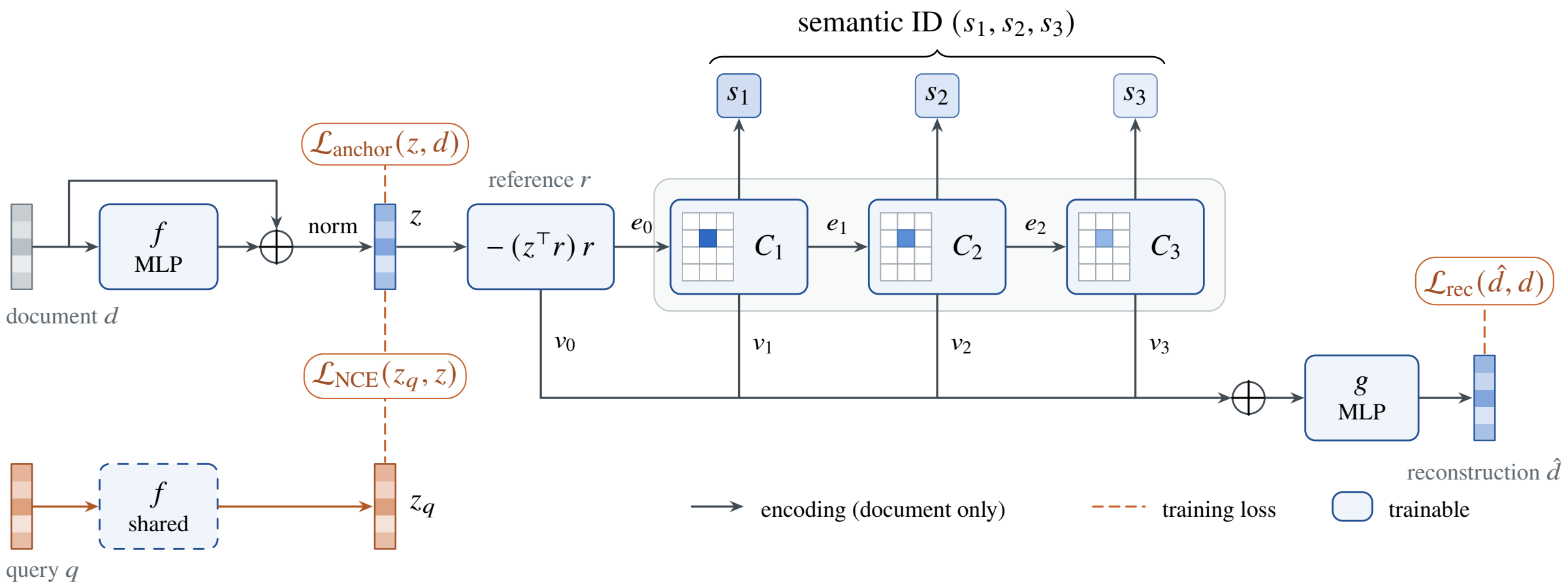

The document embedding comes from an upstream retrieval model, and quantizing it directly assumes that this space is already the best one to partition. CARE-SID instead quantizes in a learned space (Figure~\ref{fig:care-sid}). A residual alignment network in front of the quantizer is trained with a query--document contrastive loss so that the quantized space moves toward what retrieval needs, while reconstruction and anchoring keep the space from drifting away from the original embedding or collapsing. Quantization uses the reference vector and rating residual quantization of R3-VAE~\citep{wan2026r3vae} with hard top-1 cosine code selection, as in its public implementation.\footnote{\url{https://github.com/wwqq/R3-VAE}}

\subsection{Model}
\label{sec:care-model}

\paragraph{Residual alignment network.}
The normalized document embedding passes through a two-layer MLP and is added back as a residual,
\begin{equation}
  \begin{gathered}
  z=\vnorm\bigl(d+f(d)\bigr),\\
  f(d)=W_2\,\GELU(W_1d+b_1)+b_2,
  \end{gathered}
\end{equation}
with a hidden width of 1024 and $\vnorm(x)=x/\lVert x\rVert$. The output layer $(W_2,b_2)$ is initialized to zero, so $z=d$ at the start of training. Codebooks fitted earlier on $d$ therefore remain valid when this stage begins, and the space changes gradually as training proceeds. During training, the same network maps each query embedding $q$ to $z_q=\vnorm(q+f(q))$.

\paragraph{Reference removal and rating residual quantization.}
A learned unit reference vector $r$ absorbs a direction shared by all documents. Its component is removed, and the remainder is quantized over three levels. At each level the residual is normalized, the codeword with the largest cosine is selected, and the projection of the residual onto that codeword is subtracted:
\begin{equation}
\begin{aligned}
  v_0 &= (z^{\top}r)\,r,\quad e_0 = z-v_0,\\
  \bar e_{\ell-1} &= \vnorm(e_{\ell-1}),\quad s_\ell = \argmax_k\, \bar e_{\ell-1}^{\top}c_{\ell,k},\\
  \rho_\ell &= \bar e_{\ell-1}^{\top}c_{\ell,s_\ell},\quad v_\ell = \rho_\ell\,c_{\ell,s_\ell},\\
  e_\ell &= \bar e_{\ell-1}-v_\ell,\qquad \ell=1,2,3,
\end{aligned}
\label{eq:rating}
\end{equation}
where the codewords $c_{\ell,k}$ are the normalized rows of the codebook $C_\ell$, and $\rho_\ell$, the cosine between the normalized residual and the selected codeword, is called the rating. Unlike $k$-means, which subtracts a centroid, rating quantization subtracts a projection and operates on normalized residuals at every level. The reference direction does not occupy an SID position. The output $\rho_\ell c_{\ell,s_\ell}$ is differentiable with respect to the selected codeword and to the input, so training needs no straight-through estimator. The SID is $(s_1,s_2,s_3)$; the rating and reference coefficients vary with the document, are not part of the SID, and allow two documents with the same SID to have different reconstructions.

\paragraph{Reconstruction head.}
The reference and rating components are summed and decoded by a $256\to1024\to256$ MLP $g$ with ReLU,
\begin{equation}
  \dhat=\vnorm\bigl(g(t)\bigr),\quad t=v_0+v_1+v_2+v_3.
\end{equation}
Encoding a document into its SID needs only the alignment network, the reference vector and the three codebooks (Algorithm~\ref{alg:encode}); the head is used only for training and for the reconstruction view.

\begin{algorithm*}[t]
\caption{CARE-SID encoding of one document (document-only)}
\label{alg:encode}
\begin{algorithmic}[1]
\Require embedding $d$; alignment network $f$; reference $r$; codebooks $C_1,C_2,C_3$ with normalized rows $c_{\ell,k}$
\Ensure SID $(s_1,s_2,s_3)$ and components $v_0,\dots,v_3$
\State $z\gets\vnorm(d+f(d))$;\quad $v_0\gets(z^{\top}r)\,r$;\quad $e_0\gets z-v_0$
\For{$\ell=1,2,3$}
  \State $\bar e_{\ell-1}\gets\vnorm(e_{\ell-1})$;\quad $s_\ell\gets\argmax_k \bar e_{\ell-1}^{\top}c_{\ell,k}$
  \State $\rho_\ell\gets \bar e_{\ell-1}^{\top}c_{\ell,s_\ell}$;\quad $v_\ell\gets\rho_\ell\,c_{\ell,s_\ell}$;\quad $e_\ell\gets \bar e_{\ell-1}-v_\ell$
\EndFor
\State \Return $(s_1,s_2,s_3)$ and $(v_0,v_1,v_2,v_3)$ \Comment{$\dhat=\vnorm(g(\textstyle\sum_{\ell=0}^{3}v_\ell))$ if needed}
\end{algorithmic}
\end{algorithm*}

\subsection{Training objective}
\label{sec:care-loss}

The reconstruction, anchor, consistency and co-cluster terms are computed on batches of documents; the contrastive term is computed on batches of positive query--document pairs:
\begin{equation}
\begin{aligned}
  \mathcal{L} &= \Lrec+\beta\,\Lanc+\gamma\,\Lnce\\
  &\quad+\lambda_1\Lsc+\lambda_2\Lcl,\\
  \Lrec &= 1-\cos(d,\dhat),\quad \Lanc=1-\cos(z,d),\\
  \Lnce &= -\frac{1}{2B}\sum_{i=1}^{B}\Bigl[\log\frac{e^{\sigma_{ii}/\tau}}{\sum_j e^{\sigma_{ij}/\tau}}\\
  &\qquad\qquad\quad+\log\frac{e^{\sigma_{ii}/\tau}}{\sum_j e^{\sigma_{ji}/\tau}}\Bigr],\\
  \sigma_{ij} &= \cos(z_{q_i},z_{d_j}).
\end{aligned}
\label{eq:care-loss}
\end{equation}
The reconstruction target is the original embedding $d$, not the aligned $z$, so the alignment network cannot lower the reconstruction loss by collapsing the aligned space. The anchor loss keeps $z$ close to $d$. The contrastive loss is a bidirectional InfoNCE~\citep{oord2018cpc} over $B$ positive pairs, with the other documents and queries in the batch as negatives, applied to the aligned vectors \emph{before} quantization. The consistency loss follows the R3-VAE implementation and compares the reference-removed input with the cumulative quantization,
\[
  \Lsc=\frac{1}{3}\sum_{\ell=1}^{3}\Bigl[1-\cos\Bigl(e_0,\sum_{\ell'=0}^{\ell}v_{\ell'}\Bigr)\Bigr].
\]
The soft co-cluster loss, a differentiable first-level relaxation of the semantic-cohesion measure of R3-VAE~\citep{wan2026r3vae}, encourages documents whose original embeddings are similar to share a first-level bucket: with soft assignments $w_i=\softmax_k\bigl(\cos(e_{0,i},c_{1,k})/\tau_c\bigr)$ and co-assignment weights $P_{ij}=w_i^{\top}w_j$ in a batch,
\begin{equation}
  \Lcl = 1-\frac{\sum_{i\neq j}P_{ij}\cos(d_i,d_j)}{\sum_{i\neq j}P_{ij}},
\end{equation}
where $e_0$ is treated as a constant, so this term updates only the first-level codebook. We use $\beta=1.0$, $\gamma=0.03$, $\lambda_1=0.1$, $\lambda_2=0.01$, $\tau=0.1$ and $\tau_c=0.07$, with 2,048 documents and 8,192 positive pairs per step.

\subsection{Staged warm start}
\label{sec:care-stages}

CARE-SID is trained in three stages, each initialized from the previous one. Stage~1 fits RQ-KMeans codebooks on the corpus. Stage~2 trains an R3-VAE (reference vector, rating codebooks and reconstruction head, with $\Lrec+\lambda_1\Lsc+\lambda_2\Lcl$) starting from the $k$-means codebooks. Stage~3 adds the zero-initialized alignment network together with the anchor and contrastive losses and trains the full objective of Eq.~\eqref{eq:care-loss}. Because $z=d$ at initialization, Stage~3 begins exactly at the Stage~2 model. Stages~2 and~3 are each trained for 20 epochs with AdamW (learning rate $5\times10^{-4}$, weight decay $10^{-5}$; Table~\ref{tab:hparams} in Appendix~\ref{app:hparams}).

\subsection{Design analysis}
\label{sec:care-design}

Each of the following choices is backed by a controlled comparison in an earlier development round, which used a different document embedding, smaller codebooks (1024, 512 and 128 codewords per level) and, apart from the warm-start comparison, no reference vector. These comparisons motivate the design; they are not results in the setting of Section~\ref{sec:exp}.
\begin{itemize}[leftmargin=1.3em]
  \item \textbf{Warm start.} Randomly initialized codebooks collapse quickly, with ICR around 0.15; the staged warm start avoids this.
  \item \textbf{Contrast before quantization.} In a variant without the alignment network, a contrastive loss on the quantized output can move only the codebooks, and it pushed many documents into the same buckets: ICR fell from 0.856 to 0.544. Applied to $z$ before quantization and combined with the anchor loss, the contrastive loss does not cause this. WeGR~\citep{han2026wegr} likewise applies its contrastive objectives before quantization, so as not to force dissimilar items into the same SID.
  \item \textbf{Strong anchoring.} Lowering $\beta$ from 1.0 to 0.1 let the alignment network move the space too far: ICR fell from 0.850 to 0.794 and retrieval metrics dropped with it.
  \item \textbf{A small co-cluster weight.} Raising $\lambda_2$ from 0.01 to 0.1 merged too many buckets and hurt overall results.
\end{itemize}
Beyond these design choices, all query and document embeddings used in training and evaluation must share one embedding space with the Stage~1 codebooks. A mismatch is silent, because dimensions, data fields and loss curves all look normal, so before training and evaluation we compute the mean cosine between the embeddings and their reconstructions by these codebooks and abort if it is below 0.70.

\section{CARE-LR: Preserving the Ranking after Quantization}
\label{sec:carelr}

\subsection{Motivation}

\begin{figure*}[t]
  \centering
  \includegraphics[width=\textwidth]{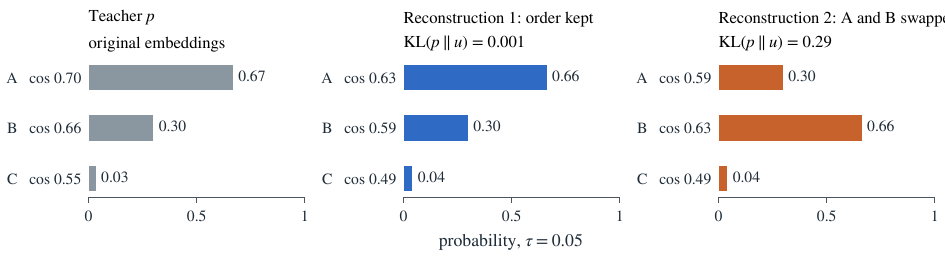}
  \caption{Same reconstruction error, different ranking. Bars are softmax probabilities over three candidates A, B and C for one query, computed from the listed cosines with $\tau=0.05$: the teacher $p$ uses the original embeddings and the student $u$ a reconstruction. Both reconstructions can have the same per-document error, but reconstruction 2 swaps A and B. A per-document loss cannot tell them apart; the listwise KL can.}
  \label{fig:kl}
\end{figure*}

The reconstruction loss of CARE-SID is computed document by document: each document is compared only with its own reconstruction. Retrieval, however, depends on the relative order of several candidates for the same query. Two documents that each move only slightly can swap a decisive pair, and two reconstructions with the same per-document error can affect the ranking very differently (Figure~\ref{fig:kl}). We also tried to constrain the discrete codes directly, for example by asking relevant documents of a query to share prefixes; this easily over-merges buckets, and in one extreme setting the number of first-level codewords in use dropped from 16,305 to 9,069 while retrieval became worse. CARE-LR therefore supervises the reconstruction pathway: it constrains the scores between a query and the reconstructions and keeps the document-only encoding interface unchanged.

\subsection{Listwise distribution distillation}

Each training query comes with a list of candidate documents. Up to 32 candidates are drawn from real search-funnel lists by round-robin over five source categories: two from the fine-ranking stage (documents labeled positive, and the remaining ones), documents from the top and from the bottom of the coarse-ranking list, and randomly sampled documents. We then add up to 16 approximate nearest neighbors of the query in the original embedding space and up to 16 random documents, both drawn from a pool of background documents sampled uniformly from the corpus, and remove duplicate document IDs. On this list, the teacher distribution is given by the query's cosines with the original embeddings and the student distribution by its cosines with the reconstructions:
\begin{equation}
\begin{aligned}
  p_j &= \softmax_j\bigl(\cos(q,d_j)/\tau\bigr),\\
  u_j &= \softmax_j\bigl(\cos(q,\dhat_j)/\tau\bigr),\\
  \Lkl &= \mathbb{E}_q\sum_j p_j\log\frac{p_j}{u_j},\\
  \Lrec &= \mathbb{E}_d\bigl[1-\cos(d,\dhat)\bigr],\\
  \mathcal{L} &= \Lrec+0.1\,\Lkl,\qquad \tau=0.05.
\end{aligned}
\label{eq:lr-loss}
\end{equation}
The teacher depends only on cosines of the original embeddings. List positions, ranker scores, relevance grades and clicks are never used as targets; the candidate sources only stratify sampling, so that each list contains both strong competitors and easy documents. Because the KL is normalized within a list, it constrains the relative score gaps between candidates rather than each cosine separately. The reconstruction loss is computed on background documents and keeps the reconstruction from degrading over the document distribution as a whole.

\subsection{Layer-resolved reconstruction head}

The CARE-SID head adds the four components before mapping them, so the corrections made by the second and third levels cannot be told apart once summed. CARE-LR keeps the summed path and adds three maps of component differences:
\begin{equation}
\begin{aligned}
  \text{CARE-SID:}\quad h &= Wt+b,\\
  \text{CARE-LR:}\quad h &= Wt+A_0(v_0-v_1)\\
  &\quad+A_2(v_2-v_1)\\
  &\quad+A_3(v_3-v_1)+b,\\
  \text{both:}\quad \dhat &= \vnorm\bigl(W_o\ReLU(h)+b_o\bigr),
\end{aligned}
\label{eq:layer-head}
\end{equation}
where $t=v_0+v_1+v_2+v_3$, $(W,b)$ and $(W_o,b_o)$ are the two layers of the CARE-SID head $g$, and $W,A_0,A_2,A_3\in\mathbb{R}^{1024\times256}$. This is equivalent to giving each component its own first-layer matrix ($W+A_0$ for $v_0$, $W+A_2$ and $W+A_3$ for $v_2,v_3$, and $W-A_0-A_2-A_3$ for $v_1$). The three new maps are initialized to zero and all other parameters are copied from CARE-SID, so that, as with the function-preserving transformations of Net2Net~\citep{chen2016net2net}, training starts from exactly the CARE-SID function. The new maps add 786,432 parameters. For example, two documents with the same reference component and rating components $(v_1,v_2,v_3)$ and $(v_1,v_2+\delta,v_3-\delta)$ have the same sum $t$ and are indistinguishable to the summed head, whereas the layer-resolved head maps them to first-layer pre-activations $h$ that differ by $(A_2-A_3)\delta$ and can therefore tell whether a correction came from the second or the third level.

\subsection{Training}

\begin{figure*}[t]
  \centering
  \resizebox{\textwidth}{!}{
\begin{tikzpicture}[x=1cm, y=1cm]
  \begin{scope}[shift={(0,4.25)}]
    \node[font=\small\bfseries, anchor=north west] at (-0.05,3.55) {(a)};
    \begin{scope}[shift={(10.0,3.33)}, font=\scriptsize, every node/.append style={text height=1.5ex, text depth=.25ex}]
      \node[frozen, minimum height=0.3cm, minimum width=0.42cm, inner sep=0pt] at (0,0) {};
      \node[anchor=west] at (0.25,0) {frozen \textcolor{cGray!90!black}{\faSnowflake[regular]}};
      \node[module, minimum height=0.3cm, minimum width=0.42cm, inner sep=0pt] at (1.95,0) {};
      \node[anchor=west] at (2.2,0) {trainable};
      \node[newmod, minimum height=0.3cm, minimum width=0.42cm, inner sep=0pt] at (3.75,0) {};
      \node[anchor=west] at (4.0,0) {new, zero-initialized};
    \end{scope}

    \pic at (0.3,1.35) {vec=cGray};
    \node[note, anchor=north west] at (0.0,0.85) {document $d$};
    \node[frozen, minimum width=2.0cm] (enc) at (2.15,1.35)
      {$f,\ r,\ C_1$\\[-1pt]{\scriptsize from CARE-SID \textcolor{cGray!90!black}{\faSnowflake[regular]}}};
    \node[module, minimum width=1.45cm, minimum height=1.0cm] (c2) at (4.75,1.35) {};
    \pic at (4.42,1.35) {codebook=cL2};
    \node[font=\small] at (5.08,1.35) {$C_2$};
    \node[module, minimum width=1.45cm, minimum height=1.0cm] (c3) at (6.75,1.35) {};
    \pic at (6.42,1.35) {codebook=cL3};
    \node[font=\small] at (7.08,1.35) {$C_3$};
    \draw[flow] (0.42,1.35) -- (enc.west);
    \draw[flow] (enc.east) -- node[above, font=\scriptsize] {$e_1$} (c2.west);
    \draw[flow] (c2.east) -- node[above, font=\scriptsize] {$e_2$} (c3.west);
    \foreach \n/\x/\c/\l in {enc/2.15/cL1/1, c2/4.75/cL2/2, c3/6.75/cL3/3} {
      \node[tok=\c] (t\l) at (\x,2.75) {$s_{\l}$};
      \draw[flow] (\n.north) -- (t\l.south);
    }
    \draw[decorate, decoration={brace, amplitude=4pt}, line width=0.5pt]
      ([xshift=-2pt, yshift=3pt]t1.north west) -- ([xshift=2pt, yshift=3pt]t3.north east)
      node[midway, above=4pt, font=\small] {semantic ID $(s_1,s_2,s_3)$, unchanged interface};

    \coordinate (bus) at (0,0.25);
    \foreach \n/\lab in {enc/{v_0,v_1}, c2/{v_2}, c3/{v_3}}
      \draw[line width=0.65pt, draw=cInk!85] (\n.south) -- node[right, font=\scriptsize] {$\lab$} (\n.south |- bus);
    \draw[flow] (enc.south |- bus) -- (8.55,0.25);

    \begin{scope}[on background layer]
      \node[group, fit={(8.55,-0.1) (13.2,2.25)}, inner sep=0pt] (head) {};
    \end{scope}
    \node[note, anchor=south west] at (8.6,2.27) {layer-resolved reconstruction head};
    \draw[line width=0.65pt, draw=cInk!85] (8.55,0.25) -- (8.8,0.25);
    \draw[line width=0.65pt, draw=cInk!85] (8.8,0.25) -- (8.8,1.95);
    \foreach \y/\inp/\mapname/\sty [count=\k] in {1.95/{t=\textstyle\sum_\ell v_\ell}/W/module, 1.4/{v_0-v_1}/A_0/newmod,
                                   0.85/{v_2-v_1}/A_2/newmod, 0.3/{v_3-v_1}/A_3/newmod} {
      \node[font=\scriptsize, anchor=west] at (8.88,\y) {$\inp$};
      \node[\sty, minimum height=0.42cm, minimum width=0.7cm, inner sep=1pt, font=\scriptsize] (b\k) at (11.15,\y) {$\mapname$};
      \draw[flow] (10.45,\y) -- (b\k.west);
      \draw[line width=0.6pt, draw=cInk!85] (b\k.east) -- (11.85,\y);
    }
    \draw[line width=0.6pt, draw=cInk!85] (11.85,0.3) -- (11.85,1.95);
    \node[addnode] (hsum) at (12.35,1.12) {};
    \draw[line width=0.6pt] (hsum.north) -- (hsum.south) (hsum.west) -- (hsum.east);
    \draw[flow] (11.85,1.12) -- (hsum.west);
    \node[font=\scriptsize] at (12.35,1.52) {$+\,b$};
    \node[module, minimum width=1.1cm, minimum height=0.8cm, font=\scriptsize, align=center] (out) at (14.0,1.12)
      {ReLU,\\$W_o$, norm};
    \draw[flow] (hsum.east) -- node[above, font=\scriptsize] {$h$} (out.west);
    \pic at (15.55,1.12) {vec=cBlue};
    \node[note, below=0.5cm] at (15.55,1.12) {reconstruction $\hat d$};
    \draw[flow] (out.east) -- (15.42,1.12);
  \end{scope}

  \draw[cGray!40, line width=0.4pt] (-0.05,4.0) -- (16.5,4.0);
  \node[font=\small\bfseries, anchor=north west] at (-0.05,3.92) {(b)};

  \pic at (0.3,1.9) {vec=cOrange};
  \node[note, anchor=north west] at (0.0,1.4) {query $q$};
  \foreach \i/\c in {0/cGray!65, 1/cGray!65, 2/cGray!65, 3/cGray!65, 4/cBlue!45, 5/cBlue!45, 6/white, 7/white} {
    \draw[cGray!90, fill=white, rounded corners=1pt, line width=0.4pt] (1.3,{3.0-0.31*\i}) rectangle ++(1.25,0.24);
    \fill[\c] (1.32,{3.02-0.31*\i}) rectangle ++(0.16,0.2);
    \foreach \k in {0,1,2} \draw[cGray!70, line width=0.35pt] ({1.6+0.25*\k},{3.12-0.31*\i}) -- ++(0.18,0);
  }
  \node[note, anchor=south] at (1.92,3.27) {candidates $d_1,\dots,d_n$};
  \draw[flow, draw=cOrange!90!black] (0.42,1.9) -- (1.28,1.9);

  \draw[line width=0.65pt, draw=cInk!85] (2.6,1.9) -- (3.1,1.9);
  \draw[line width=0.65pt, draw=cInk!85] (3.1,1.0) -- (3.1,2.7);
  \node[frozen, minimum width=2.7cm, minimum height=0.75cm, font=\scriptsize] (tea) at (5.15,2.7)
    {$\operatorname{softmax}_j\!\big(\!\cos(q,d_j)/\tau\big)$};
  \node[note, above=0pt] at (tea.north) {teacher: original embeddings \textcolor{cGray!90!black}{\faSnowflake[regular]}};
  \node[module, minimum width=2.7cm, minimum height=0.75cm, font=\scriptsize] (stu) at (5.15,1.0)
    {$\operatorname{softmax}_j\!\big(\!\cos(q,\hat d_j)/\tau\big)$};
  \node[note, below=0pt] at (stu.south) {student: reconstructions from (a)};
  \draw[flow] (3.1,2.7) -- (tea.west);
  \draw[flow] (3.1,1.0) -- (stu.west);
  \foreach \h [count=\k] in {0.62,0.34,0.17,0.09,0.05}
    \fill[cGray!75] ({7.15+0.2*\k},2.37) rectangle ++(0.15,\h);
  \foreach \h [count=\k] in {0.56,0.38,0.18,0.10,0.06}
    \fill[cBlue!80] ({7.15+0.2*\k},0.67) rectangle ++(0.15,\h);
  \draw[cGray!70, line width=0.4pt] (7.3,2.37) -- (8.35,2.37) (7.3,0.67) -- (8.35,0.67);
  \node[font=\small] at (8.65,2.65) {$p$};
  \node[font=\small] at (8.65,0.95) {$u$};
  \draw[flow] (tea.east) -- (7.25,2.7);
  \draw[flow] (stu.east) -- (7.25,1.0);

  \node[losspill, font=\small] (kl) at (10.85,1.85) {$\mathcal{L}_{\mathrm{KL}}=\mathrm{KL}(p\,\Vert\,u)$};
  \draw[loss] (8.85,2.65) -| (kl.north);
  \draw[loss] (8.85,0.95) -| (kl.south);
  \node[draw=cOrange!85!black, line width=0.7pt, rounded corners=3pt, inner sep=4pt, font=\small] (tot) at (14.75,1.85)
    {$\mathcal{L}=\mathcal{L}_{\mathrm{rec}}+0.1\,\mathcal{L}_{\mathrm{KL}}$};
  \node[losspill, font=\small, align=center] (rec) at (14.75,0.55)
    {$\mathcal{L}_{\mathrm{rec}}=1-\cos(d,\hat d)$\\[-1pt]{\scriptsize on background documents}};
  \draw[loss, ->] (kl.east) -- (tot.west);
  \draw[loss, ->] (rec.north) -- (tot.south);

  \begin{scope}[shift={(0,-0.38)}, font=\scriptsize]
    \node[anchor=west, text=cInk!70] at (-0.05,0) {candidates:};
    \fill[cGray!65] (1.65,-0.09) rectangle ++(0.16,0.18);
    \node[anchor=west] at (1.85,0) {search-funnel lists, $\le$32 (5 categories)};
    \fill[cBlue!45] (6.8,-0.09) rectangle ++(0.16,0.18);
    \node[anchor=west] at (7.0,0) {nearest neighbors, $\le$16};
    \draw[cGray!90, line width=0.4pt] (10.35,-0.09) rectangle ++(0.16,0.18);
    \node[anchor=west] at (10.55,0) {random, $\le$16};
    \node[anchor=west, text=cInk!70] at (12.55,0) {deduplicated by document};
  \end{scope}
\end{tikzpicture}\unskip}
  \caption{CARE-LR. (a) SIDs come from the document alone: the frozen CARE-SID path ($f$, $r$, $C_1$) is followed by the trainable codebooks $C_2$ and $C_3$, so the encoding interface is unchanged. The layer-resolved head adds zero-initialized maps $A_0,A_2,A_3$ of component differences to the summed path $Wt$. (b) For each training query, the candidates (from search-funnel lists, whose five source categories only stratify sampling, plus nearest neighbors and random documents) give a teacher distribution from the original embeddings and a student one from the reconstructions; the KL aligns them, and the reconstruction loss on background documents keeps the head faithful.}
  \label{fig:care-lr}
\end{figure*}
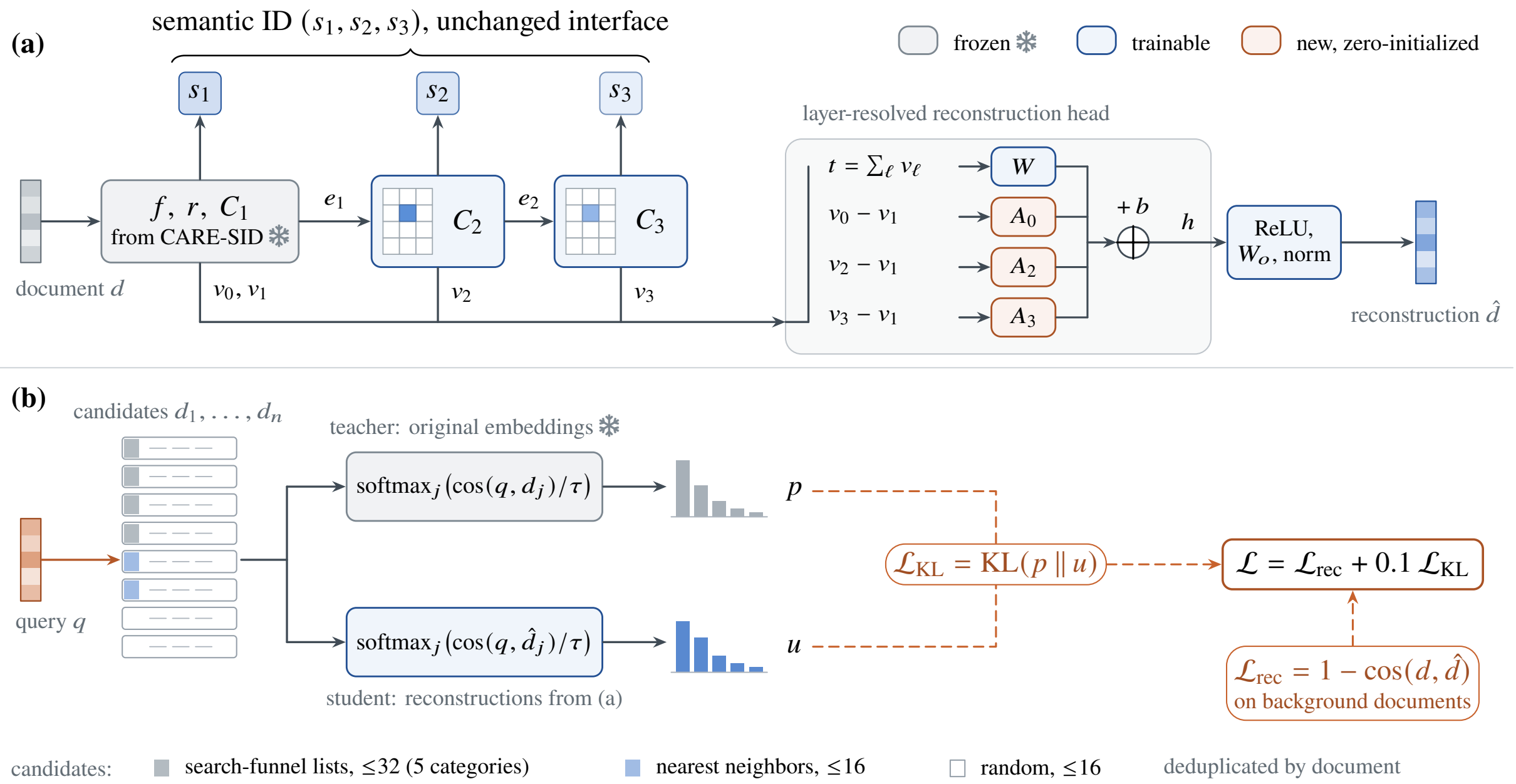

CARE-LR starts from CARE-SID and freezes the alignment network, the reference vector and the first-level codebook; it updates the second- and third-level codebooks and the reconstruction head (Figure~\ref{fig:care-lr}). Code selection remains hard top-1 at every level. Gradients reach the selected codewords through the ratings in Eq.~\eqref{eq:rating}, so when $C_2$ and $C_3$ move, the code boundaries move with them and the last two digits of some documents can change. The first digit cannot change, because everything it depends on is frozen. Each step draws a batch of background documents for $\Lrec$ and a batch of queries with their candidate lists for $\Lkl$ (Algorithm~\ref{alg:lr}). We train for 1,000 steps with Adam (learning rate $10^{-4}$, gradient norm clipped at 5). Every 100 steps, we evaluate the criterion $(1-\overline{\cos}(d,\dhat))+0.1\,\Lkl$ on 8,192 held-out background documents and 256 development queries and keep the checkpoint with the lowest criterion among those whose held-out reconstruction cosine is at most 0.001 below that of the starting model; step 0 is eligible. Held-out documents are excluded from all training lists, and neither training nor checkpoint selection uses the ideal sets or their recall. Training queries that match or closely resemble an evaluation query are removed, and candidate lists and background documents exclude the documents of the evaluation pool, the ideal documents and their near-duplicates.

\begin{algorithm*}[t]
\caption{One CARE-LR training step}
\label{alg:lr}
\begin{algorithmic}[1]
\Require CARE-SID model with frozen $f$, $r$, $C_1$; trainable $C_2$, $C_3$ and reconstruction head; $\tau=0.05$
\State Sample 512 background documents and reconstruct them with Algorithm~\ref{alg:encode} and the head
\State $\Lrec\gets\operatorname{mean}_d\bigl(1-\cos(d,\dhat)\bigr)$
\State Sample 8 training queries; give each a deduplicated candidate list of $\le32$ funnel documents
\Statex \hspace{\algorithmicindent}(round-robin over sources), $\le16$ nearest neighbors and $\le16$ random documents
\State $p\gets\softmax(\cos(q,d_j)/\tau)$ \Comment{teacher, no gradient}
\State $u\gets\softmax(\cos(q,\dhat_j)/\tau)$ \Comment{student, through hard top-1 selection and the head}
\State $\Lkl\gets\operatorname{mean}_q\KL(p\,\Vert\,u)$;\quad $\mathcal{L}\gets\Lrec+0.1\,\Lkl$
\State Clip the gradient norm at 5 and take an Adam step on $C_2$, $C_3$ and the head
\end{algorithmic}
\end{algorithm*}

\section{Experiments}
\label{sec:exp}

The experiments ask how much CARE-SID and CARE-LR improve retrieval over RQ-KMeans and R3-VAE, where the gain of CARE-LR over CARE-SID comes from and whether it changes the partition that a GR model consumes, and how the gain of CARE-SID over R3-VAE depends on the document embedding.

\subsection{Setup}

\paragraph{Data.}
The models are trained on production-scale document embeddings and positive query--document pairs from our video search service. The evaluation pool, shared by all methods, contains 1,001,700 document rows, one per document (so $c=1$ in Eq.~\eqref{eq:pi}): documents sampled at random from the service, to which the ideal documents of the evaluation queries were added. The queries form four segments. In the general and long-tail segments, the ideal documents were annotated manually; in the high-click and high-exposure segments, which consist of queries with many clicks or exposures over seven days, they are derived from user behavior. Queries without an available embedding are dropped, which leaves 704, 292, 995 and 999 evaluable queries (2,990 in total); results are averaged over all segment--query pairs, so each segment is weighted by its number of queries. The few ideal documents without an available embedding are absent from the pool but remain in the denominator of recall, so no method can reach 100\%. Every method encodes the pool with its own tokenizer; queries keep their original embeddings, are neither passed through the alignment network nor quantized, and are compared with the bucket keys by cosine similarity.

\paragraph{Methods.}
RQ-KMeans and R3-VAE are the baselines, and CARE-SID and CARE-LR are our methods. The two baselines are the models after Stages~1 and~2 of the staged warm start of CARE-SID (Section~\ref{sec:care-stages}), so the difference between R3-VAE and CARE-SID measures Stage~3 as a whole: the alignment network, the anchor and contrastive losses, and 20 further epochs of training. The CARE-LR ablations are initialized from the same CARE-SID checkpoint and trained for 1,000 steps with the same data order, candidates and selection rule. All results in Table~\ref{tab:main} except R3-VAE come from the same evaluation code, pool and queries; the two arms without the KL were evaluated in a second run. R3-VAE comes from an earlier evaluation on the same pool and queries, which drew rows inside the partially consumed bucket once at random instead of using Eq.~\eqref{eq:pi} and recorded recall to four decimal places. The same CARE-SID model agrees between the two evaluations to within 0.01 points at Recall@10, @50 and @100.

\paragraph{Scale of the CARE-LR experiments.}
CARE-SID is trained at production scale, whereas the CARE-LR results in this version come from a small-scale pilot: its training queries and candidate lists are drawn from about one million search-funnel lists, and each run lasts 1,000 steps with 8 queries and 512 background documents per step. The CARE-LR numbers below are therefore a first validation of the method. We will update them with larger-scale training in a later version.

\subsection{Main results}

\begin{table*}[t]
\centering
\small
\setlength{\tabcolsep}{4.5pt}
\caption{Reconstruction-view Recall@$N$ (\%, \view{recon\_P3}). Our methods are in bold. Paired differences are in points and computed before rounding; bold differences have a 95\% CI that excludes zero, and $^{\circ}$ marks a CI that includes zero. $^\dagger$From an earlier evaluation on the same pool and queries (Section~\ref{sec:exp}, Methods).}
\label{tab:main}
\begin{tabular}{llrrrrrr}
\toprule
Method & Description & @5 & @10 & @20 & @30 & @50 & @100 \\
\midrule
\multicolumn{8}{l}{\textit{Baselines}} \\
RQ-KMeans & three-level residual $k$-means & 53.71 & 62.75 & 69.70 & 73.14 & 77.34 & 82.11 \\
R3-VAE$^\dagger$ & reference + rating residual quant. & 56.02 & 65.15 & 72.20 & 75.67 & 79.58 & 83.91 \\
\midrule
\multicolumn{8}{l}{\textit{Ours}} \\
\textbf{CARE-SID} & alignment net + anchored contrastive & 56.09 & 65.23 & 72.35 & 75.92 & 80.24 & 84.43 \\
\textbf{CARE-LR} & CARE-SID + listwise reconstruction & 56.72 & 66.09 & 72.88 & 76.30 & 80.46 & 84.74 \\
\quad CARE-LR $-$ RQ-KMeans & paired difference (points) & \textbf{+3.01} & \textbf{+3.34} & \textbf{+3.18} & \textbf{+3.15} & \textbf{+3.12} & \textbf{+2.62} \\
\quad CARE-LR $-$ CARE-SID & paired difference (points) & \textbf{+0.63} & \textbf{+0.86} & \textbf{+0.53} & \textbf{+0.37} & +0.21$^{\circ}$ & \textbf{+0.31} \\
\midrule
\multicolumn{8}{l}{\textit{CARE-LR ablations (each initialized from CARE-SID and trained for 1{,}000 steps)}} \\
Summed head, no KL & without KL and layer-resolved head & 56.14 & 65.28 & 72.39 & 75.89 & 80.18 & 84.41 \\
Summed head + KL & without layer-resolved head & 56.64 & 65.97 & 72.80 & 76.23 & 80.48 & 84.66 \\
Layer-resolved head, no KL & without KL & 56.11 & 65.24 & 72.39 & 75.92 & 80.18 & 84.41 \\
Frozen codebooks & head only; SIDs unchanged & 56.74 & 66.12 & 72.88 & 76.35 & 80.46 & 84.72 \\
\bottomrule
\end{tabular}

\end{table*}

\begin{figure*}[t]
  \centering
  \includegraphics[width=\textwidth]{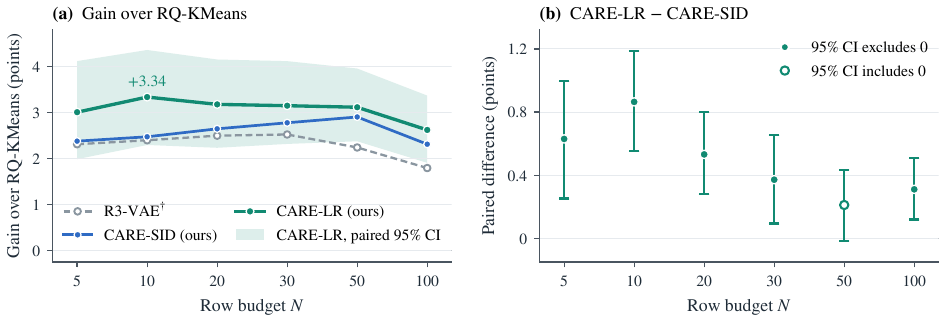}
  \caption{Reconstruction-view Recall@$N$. (a) Gains of R3-VAE, CARE-SID and CARE-LR over RQ-KMeans; the band is the paired 95\% CI of CARE-LR (R3-VAE$^\dagger$, from the earlier evaluation, has no paired interval). (b) Paired difference between CARE-LR and CARE-SID with 95\% CIs; filled markers denote intervals that exclude zero.}
  \label{fig:results}
\end{figure*}

Table~\ref{tab:main} and Figure~\ref{fig:results} show that both of our methods improve on RQ-KMeans by more than two points at every reported cutoff, and that CARE-LR also improves on R3-VAE. CARE-SID reaches 65.23\% \view{recon\_P3}@10, 2.48 points above RQ-KMeans. CARE-LR reaches 66.09\%, 3.34 points above RQ-KMeans (95\% CI [+2.30, +4.36]), and the paired interval of its gain over RQ-KMeans excludes zero at every reported cutoff. R3-VAE, the quantization kernel of CARE-SID, is close to CARE-SID at @10 and 0.5--0.7 points lower at @50 and @100; with listwise distillation, CARE-LR exceeds R3-VAE by 0.94 points at @10. Section~\ref{sec:exp-emb} shows that the gain of CARE-SID over R3-VAE is considerably larger when the document embedding has not been trained on search data.

CARE-LR improves on CARE-SID by 0.86 points at @10 ([+0.55, +1.19]) and by 0.31 points at @100 ([+0.12, +0.51]); the improvement is concentrated in the head of the ranking. The row-level reconstruction view gives the same conclusion: \view{recon\_doc}@10 rises from 65.22\% to 66.10\%.

\subsection{Ablations: where the gain of CARE-LR comes from}

\begin{figure*}[t]
  \centering
  \includegraphics[width=\textwidth]{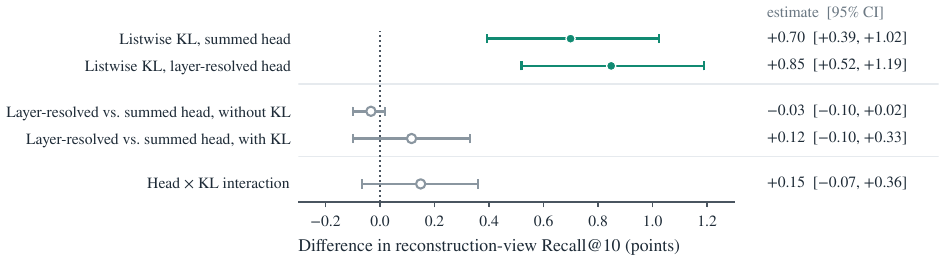}
  \caption{Effects on \view{recon\_P3}@10 with paired 95\% CIs. Rows 1--2: adding the listwise KL to each head; rows 3--4: layer-resolved minus summed head; row 5: the head $\times$ KL interaction, bootstrapped directly on the per-query double differences of the $2\times2$ design. Filled markers denote intervals that exclude zero.}
  \label{fig:ablation}
\end{figure*}

CARE-LR adds the listwise KL and the layer-resolved head and lets the second- and third-level codebooks move. The lower block of Table~\ref{tab:main} removes these changes singly and in combination: its first three rows form, together with CARE-LR, a $2\times2$ design (head structure $\times$ use of the KL) whose effects are shown in Figure~\ref{fig:ablation}, and its last row freezes the codebooks. In this design, both arms trained with the reconstruction loss alone were selected at step 200 and both arms trained with the KL at step 300, so no arm was selected at its starting point. Adding the KL raises \view{recon\_P3}@10 from 65.28\% to 65.97\% (+0.70, [+0.39, +1.02]) with the summed head and from 65.24\% to 66.09\% (+0.85, [+0.52, +1.19]) with the layer-resolved head; for both heads the intervals exclude zero at every cutoff from $N=1$ to $N=100$. The head structure on its own matters less: with the KL, the layer-resolved head is 0.12 points above the summed head ([$-$0.10, +0.33]), and the interaction is +0.15 points ([$-$0.07, +0.36]), both with intervals that include zero. \textbf{The gain of CARE-LR over CARE-SID comes mainly from the listwise distillation.}

The frozen-codebook arm trains only the layer-resolved head, with the same objective and budget as CARE-LR, so every document keeps the SID assigned by CARE-SID; CARE-LR minus this arm isolates the effect of updating the codebooks. It reaches 66.12\%, on par with the full CARE-LR (difference $-$0.03, [$-$0.12, +0.05]). In this pilot, the improvement is therefore realized through the reconstruction pathway, and updating the codebooks brought no additional measurable gain.

\subsection{Bucket structure}

The bucketing views depend only on the partition induced by the integer SIDs, which is what a GR model consumes. The reconstruction-view gains of CARE-LR leave the bucketing quality of this partition nearly unchanged. Because CARE-LR freezes the alignment network, the reference vector and the first-level codebook, its one-digit buckets are identical to those of CARE-SID, and at two- and three-digit prefixes the bucketing-view recall of the two models differs by at most 0.02 points at every cutoff up to $N=100$. Downstream models that consume integer SIDs therefore see a prefix tree of essentially the same bucketing quality, although the last two digits of some documents change (Section~\ref{sec:deploy}). Compared with RQ-KMeans, the jointly evaluated baseline, the one-digit bucketing of CARE-SID is better, with \view{rawmean\_P1}@100 of 50.94\% versus 48.43\% (+2.51 points), and at two-digit prefixes the two are close, with RQ-KMeans ahead by 0.1 to 0.8 points.

\subsection{Effect of the document embedding}
\label{sec:exp-emb}

\begin{figure*}[t]
  \centering
  \includegraphics[width=\textwidth]{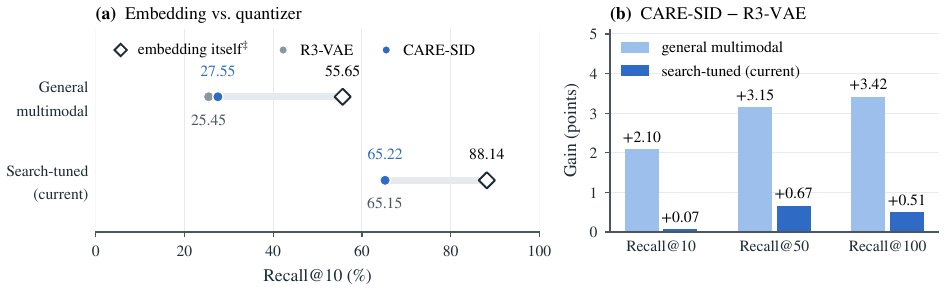}
  \caption{Two document embeddings under the same pool, queries, codebook sizes and hyperparameters. (a) Recall@10 of the embedding itself ($^\ddagger$\view{rawmean\_P3}, see Table~\ref{tab:emb}) and reconstruction-view Recall@10 of R3-VAE and CARE-SID: changing the embedding moves recall far more than changing the quantizer. (b) Gain of CARE-SID over R3-VAE in reconstruction-view recall (points). Both panels use the earlier evaluation protocol (Table~\ref{tab:emb}); for the search-tuned embedding the R3-VAE and CARE-SID markers in (a) overlap.}
  \label{fig:emb}
\end{figure*}

How much the contrastive alignment of CARE-SID can add depends strongly on the document embedding. Our first experiments used a general multimodal embedding that had not been trained on search query--document data; the current system uses a search-tuned retrieval embedding, trained contrastively on search data of the same service. We trained R3-VAE and CARE-SID on each embedding with the same codebook sizes and hyperparameters, on training documents and query--document pairs sampled separately for each embedding, and evaluated them on the same document pool, queries and ideal sets (Table~\ref{tab:emb}, Figure~\ref{fig:emb}).

\begin{table*}[t]
\centering
\small
\setlength{\tabcolsep}{5pt}
\caption{R3-VAE and CARE-SID on two document embeddings (\view{recon\_P3}, \%; gains in points). Both rows come from the earlier evaluation protocol (Section~\ref{sec:exp}). The $q$--$d$ gap is the mean cosine of positive pairs (the evaluation queries and their ideal documents) minus that of random query--document pairs. Gains have no confidence intervals. $^\ddagger$\view{rawmean\_P3}@10: buckets by full SID, keyed by the mean original embedding; with about one document per bucket, this approximates retrieval with the embedding itself.}
\label{tab:emb}
\begin{tabular}{lrrrrrrr}
\toprule
& & \multicolumn{3}{c}{Recall@10 (\%)} & \multicolumn{3}{c}{CARE-SID $-$ R3-VAE} \\
\cmidrule(lr){3-5}\cmidrule(lr){6-8}
Document embedding & $q$--$d$ gap & raw$^\ddagger$ & R3-VAE & CARE-SID & @10 & @50 & @100 \\
\midrule
General multimodal & 0.27 & 55.65 & 25.45 & 27.55 & +2.10 & +3.15 & +3.42 \\
Search-tuned (current) & 0.70 & 88.14 & 65.15 & 65.22 & +0.07 & +0.67 & +0.51 \\
\bottomrule
\end{tabular}

\end{table*}

On the general multimodal embedding, the alignment stage of CARE-SID gives a clear gain: CARE-SID exceeds R3-VAE by 2.10, 3.15 and 3.42 points at Recall@10, @50 and @100. On the search-tuned embedding the gain narrows to 0.07, 0.67 and 0.51 points, and at @10 the two models are on par. We attribute the difference to how well the embedding already separates relevant from random query--document pairs. In the general embedding, a random query--document pair already has a mean cosine of 0.44, and positive pairs exceed this by only 0.27; the search-tuned embedding has been trained contrastively on the same kind of data and reaches a gap of 0.70. The upstream model has already done much of the alignment that CARE-SID's contrastive loss would add, and less signal is left for it: the InfoNCE loss starts at 2.17 instead of 4.05 and decreases by 0.08 instead of 0.30 over 20 epochs.

The choice of embedding matters far more than the choice of quantizer. Moving from the general to the search-tuned embedding raises the \view{recon\_P3}@10 of CARE-SID from 27.55\% to 65.22\% and the retrieval ability of the embedding itself from 55.65\% to 88.14\%, whereas different quantizers on the same embedding differ by a few points. In our system, improving the retrieval quality of the embedding before building SIDs was the step with the largest return. \citet{penha2025joint} observe a related effect in joint search and recommendation: semantic IDs built from search-tuned embeddings give about five times the search recall of semantic IDs built from content embeddings.

\section{Deployment in Video Search}
\label{sec:deploy}

CARE-SID is applied in the generative retrieval stage of an industrial video search system (Figure~\ref{fig:overview}). Offline, every video's document embedding is encoded into a three-digit SID with Algorithm~\ref{alg:encode}, and a mapping from SIDs to document IDs is maintained; new videos are encoded incrementally with the same weights. Encoding needs only the alignment network, the reference vector and the three codebooks; the reconstruction head is not served. Online, the GR model reads the query and the SID sequence of the user's previously clicked videos and generates the SID of a target video digit by digit,
\begin{equation}
  P(s_1,s_2,s_3\mid q,H)=\prod_{\ell=1}^{3}P(s_\ell\mid q,H,s_{<\ell}),
\end{equation}
where $H$ is the user history and each level uses its own code tokens. Decoding is constrained by a prefix tree of valid SIDs, and generated SIDs are mapped back to videos; when an SID holds several videos, they are taken in order of relevance within the bucket.

CARE-LR keeps the CARE-SID encoding interface and leaves the first digit of every video unchanged, since everything that digit depends on is frozen. Because it updates the second- and third-level codebooks, some videos receive new second or third digits, so deploying it requires refreshing the SID mapping together with the training labels and user histories of the GR model.

\section{Conclusion}
\label{sec:conclusion}

An SID tokenizer bounds what generative retrieval can recall, yet common SID constructions optimize criteria defined on document embeddings alone and leave uncontrolled where the quantization loss falls. We built and judged SID tokenizers by query--document retrieval instead, while still encoding every document from its own embedding. CARE-SID quantizes in a retrieval-aligned space: it places a zero-initialized residual alignment network, trained with anchored contrastive learning, in front of the rating residual quantizer of R3-VAE, improves reconstruction-view Recall@10 by 2.48 points over RQ-KMeans, and is deployed in the generative retrieval stage of an industrial video search system. CARE-LR distills the candidate distributions of the original embedding into the SID reconstructions; this listwise distillation accounts for most of its further gain of 0.86 points, for a total of 3.34 points over RQ-KMeans. The contrastive alignment pays off most when the embedding has not been trained on search data, and the choice of embedding moves SID recall far more than the choice of quantizer. Next, we will train CARE-LR at a larger scale, let the listwise signal reshape the discrete codes as well, so that the ranking knowledge of the embedding reaches the GR model through the integer SIDs themselves, and report end-to-end results of the GR model.

\section*{Limitations}

All results are offline and measured at the SID level. CARE-LR has so far been trained only in the small-scale pilot described in Section~\ref{sec:exp}; end-to-end results of the GR model and CARE-LR results with larger-scale training will be added in a later version. The confidence intervals reflect query sampling only, not training-seed variation, and are not corrected for multiple comparisons. The ideal sets were also used in earlier development rounds and are therefore not a fresh blind test. The reconstruction view of R3-VAE and the CARE models uses per-document rating and reference coefficients that are not part of the integer SID, so for these models it is not determined by the SID that a GR model generates. The gains of CARE-LR in this pilot are realized mainly through the reconstruction pathway, and their benefit for a GR model that consumes integer SIDs remains to be confirmed end to end. The R3-VAE baseline and the embedding comparison come from an earlier evaluation on the same pool and queries, and the design analysis of Section~\ref{sec:care-design} comes from an earlier round with a different embedding and smaller codebooks.

\bibliography{references}

\clearpage
\onecolumn  
\appendix
\section{Hyperparameters}
\label{app:hparams}

\begin{table}[h]
\centering
\small
\renewcommand{\arraystretch}{1.15}
\caption{Training configuration. All stages use $D=256$ and codebooks of sizes 16384, 4096 and 2048 with hard top-1 selection.}
\label{tab:hparams}
\setlength{\tabcolsep}{4pt}
\begin{tabular}{>{\raggedright}p{2.0cm}>{\raggedright}p{3.5cm}>{\raggedright}p{4.0cm}>{\raggedright\arraybackslash}p{5.1cm}}
\toprule
 & Stage 2: R3-VAE & Stage 3: CARE-SID & CARE-LR \\
\midrule
Initialization & RQ-KMeans codebooks (Stage 1) & Stage 2 model; zero-initialized output layer of $f$ & CARE-SID; zero-initialized $A_0,A_2,A_3$ \\
\addlinespace[2pt]
Trainable & codebooks, reference vector, reconstruction head & all of Stage 2 and the alignment network $f$ (hidden width 1024, GELU) & $C_2$, $C_3$, reconstruction head \\
\addlinespace[2pt]
Objective & $\Lrec+0.1\,\Lsc+0.01\,\Lcl$ & Stage 2 objective $+\,1.0\,\Lanc+0.03\,\Lnce$ & $\Lrec+0.1\,\Lkl$ \\
\addlinespace[2pt]
Temperatures & $\tau_c=0.07$ & $\tau=0.1$ (InfoNCE), $\tau_c=0.07$ & $\tau=0.05$ \\
\addlinespace[2pt]
Batch & 2,048 documents & 2,048 documents and 8,192 positive pairs & 512 documents and 8 queries ($\le64$ candidates each) \\
\addlinespace[2pt]
Optimizer & AdamW, lr $5\times10^{-4}$, weight decay $10^{-5}$ & AdamW, lr $5\times10^{-4}$, weight decay $10^{-5}$ & Adam, lr $10^{-4}$, no weight decay, gradient norm clipped at 5 \\
\addlinespace[2pt]
Schedule & 20 epochs & 20 epochs & 1,000 steps; validation every 100 steps \\
\addlinespace[2pt]
Checkpoint & last epoch & last epoch & $(1-\overline{\cos})+0.1\,\Lkl$ on 8,192 held-out documents and 256 development queries; held-out reconstruction cosine may drop by at most 0.001 \\
\bottomrule
\end{tabular}
\end{table}

\end{document}